\documentclass[12pt,preprint]{aastex}  
\usepackage{graphicx}
\usepackage{txfonts}
\newcommand{\cs}{c_{\mathrm{s}}}
\newcommand{\csp}{c_{\mathrm{sp}}}
\newcommand{\va}{v_{\mathrm{A}}}

\newcommand{\pd}{\partial}

\newcommand{\mutilde}{\tilde \mu}

\begin{document}

	\title{THE THERMAL INSTABILITY OF SOLAR PROMINENCE THREADS}

	\shorttitle{THE THERMAL INSTABILITY OF SOLAR PROMINENCE THREADS}

   \author{R. Soler$^{1,2}$, J. L. Ballester$^2$, and M. Goossens$^1$}

   \affil{$^1$Centre for Plasma Astrophysics, Katholieke Universiteit Leuven,
              Celestijnenlaan 200B, 3001 Leuven, Belgium}
              \email{roberto.soler@wis.kuleuven.be}
   \affil{$^2$Departament de F\'isica, Universitat de les Illes Balears,
              E-07122, Palma de Mallorca, Spain}

  \begin{abstract}

The fine structure of solar prominences and filaments appears as thin and long threads in high-resolution images. In H$\alpha$ observations of filaments, some threads can be observed for only 5 -- 20 minutes before they seem to fade and eventually disappear, suggesting that these threads may have very short lifetimes. The presence of an instability might be the cause of this quick disappearance. Here, we study the thermal instability of prominence threads as an explanation of their sudden disappearance from H$\alpha$ observations. We model a prominence thread as a magnetic tube with prominence conditions embedded in a coronal environment. We assume a variation of the physical properties in the transverse direction, so that the temperature and density continuously change from internal to external values in an inhomogeneous transitional layer representing the particular prominence-corona transition region (PCTR) of the thread. We use the nonadiabatic and resistive magnetohydrodynamic equations, which include terms due to thermal conduction parallel and perpendicular to the magnetic field, radiative losses, heating, and magnetic diffusion. We combine both analytical and numerical methods to study linear perturbations from the equilibrium state, focusing on unstable thermal solutions. We find that thermal modes are unstable in the PCTR for temperatures higher than 80,000~K, approximately. These modes are related to temperature disturbances that can lead to changes in the equilibrium due to rapid plasma heating or cooling. For typical prominence parameters, the instability time scale is of the order of a few minutes and is independent of the form of the temperature profile within the PCTR of the thread. This result indicates that thermal instability may play an important role for the short lifetimes of threads in the observations.

  \end{abstract}



\section{INTRODUCTION}

Solar prominences and filaments are large-scale magnetic structures of the solar corona. The main issues regarding the physics, dynamics, and modeling of these coronal inhabitants have been recently reviewed by \citet{labrossereview} and \citet{mackay}. High-resolution observations reveal that prominences and filaments are formed by long ($5'' - 20''$) and thin ($0''.2 - 0''.6$) fine structures, usually called threads. Although the existence of the fine structure of prominences was discovered long time ago \citep[e.g.,][]{menzel60,engvold76}, its properties and dynamics could only be studied in more detail with recent high-resolution observations. The fine structures show up as dark ribbons in H$\alpha$ images of filaments on the solar disk from the Swedish Solar Telescope \citep[e.g.,][]{lin04,lin07,lin08,lin09}, and as bright features in observations of prominences in the solar limb from the Solar Optical Telescope aboard the Hinode satellite \citep[e.g.,][]{okamoto,berger,chae,ning,brigi}. Statistical studies show that the orientation of threads with respect to the filament long axis can significantly vary within the same filament \citep{lin04}, with 20 degrees a mean value typically reported. Vertical threads are more commonly seen in quiescent prominences \citep[e.g.,][]{berger,chae} whereas horizontal threads are usually observed in active region prominences \citep[e.g.,][]{okamoto}. \citet{brigi} recently pointed out that vertical threads might actually be a pile up of horizontal threads which seem to be vertical structures when projected on the plane of the sky. However, this is still a matter of controversy. Since threads are observed in both spines and barbs, it is believed that they are the basic building blocks of prominences and filaments \citep{engvold2004}. 

Theoretically, the fine structures have been modeled as magnetic flux tubes anchored in the solar photosphere \citep[e.g.,][]{ballesterpriest,rempel}, which are only partially filled with the cool ($\sim 10^4$~K) filament material, while the rest of the tube is occupied by hot coronal plasma. Therefore, the magnetic field is oriented along the axis of the fine structure. This model is conceptually in agreement with the idea that the dense prominence material is trapped in dips near the apex of a magnetic arcade connecting two photospheric regions of opposite magnetic polarity. The dips are supposed to correspond to the observed threads, which are piled up to form the prominence body. It has also been suggested from differential emission measure studies that each thread might be surrounded by its own prominence-corona transition region (PCTR) where the plasma physical properties would abruptly vary from prominence to coronal conditions \citep{cirigliano}. 

Prominence threads are highly dynamic \citep[see, e.g.,][]{heinzel,engvold}. For example, transverse thread oscillations and propagating waves along the threads seem to be ubiquitous in prominences, which have been interpreted in terms of magnetohydrodynamic (MHD) waves \citep[see the reviews by][]{ballester, oliver, arreguiballester}. Mass flows along threads, with typical flow velocities of less than 30~km~s$^{-1}$, have been also frequently reported \citep[e.g.,][]{zirker94, zirker98, lin03, chae}. An interesting property of the observations is the apparent short lifetime of some threads when they are observed in H$\alpha$ sequences \citep[e.g.,][]{lin04,lin05,lin09}. Typically, the threads can be followed for only 5 -- 20 minutes before they seem to fade with time and eventually disappear. The cause of this quick disappearance is unknown, although several explanations have been proposed. A possible explanation is related to the presence of flows and mass motions \citep[e.g.,][]{chae,chae2,brigi}, which may trigger a Kelvin-Helmholtz instability (KHI). Recently, \citet{solerKH} investigated the KHI in magnetic flux tubes due to shear flows generated by transverse motions of the tube, while \citet{temury} studied the KHI in twisted tubes due to longitudinal flows. When the results of both papers are applied to prominences, one obtains that neither the observed transverse velocity amplitudes of oscillating threads nor the longitudinal flow velocities are large enough to trigger a KHI on time scales consistent with the apparent lasting time of the threads.  On the other hand, a different explanation was suggested by \citet{lin04} and \citet{lin05}. According to these authors, rapid cooling or heating may cause the maximum of the plasma emission to fall outside the bandpass of the filter, and so the thread would become invisible in H$\alpha$ images. A mechanism that could lead to a rapid heating or cooling of the prominence material is a thermal instability. Here, we explore this possibility by studying the thermal instability of prominence threads

Thermal or condensation modes have been extensively investigated in homogeneous plasmas \citep[e.g.,][]{parker,field,heyvaerts}. A relevant work in the context of prominences was performed by \citet{carbonell04}, who studied the thermal mode in homogeneous plasmas with prominence, PCTR, and coronal conditions, considering parallel thermal conduction to magnetic field lines and the optically thin radiative loss function of \citet{hildner}. \citet{carbonell04} obtained that, for long wavelengths, the thermal mode is unstable for PCTR temperatures since thermal conduction is not efficient enough to stabilize the thermal disturbance.  In the case of an inhomogeneous plasma, thermal modes were studied in detail in the works by \citet{vanderlindencont}, \citet{vanderlinden91} and \citet{vanderlinden93}. As the present investigation is based and inspired by these previous works, we summarize their relevant results in the following paragraph. This will also enable us to clearly define the advancement made in the present paper compared with these previous investigations.

 \citet{vanderlindencont} studied the linear nonadiabatic MHD spectrum of a magnetic cylinder taking into account the effect of plasma inhomogeneity in the transverse direction. They included radiative losses, heating, and thermal conduction parallel to the magnetic field lines but neglected perpendicular thermal conduction. These authors pointed out the important result that, along with the classical Alfv\'en and slow continua, there exists an additional thermal continuum, which can be unstable depending on the physical properties of the equilibrium. Subsequently, \citet{vanderlinden91} investigated the effect on the thermal continuum of thermal conduction perpendicular to the magnetic field lines, and showed that perpendicular thermal conduction replaces the continuum by a dense set of discrete quasi-continuum modes.  These quasi-continuum modes retain the basic stability properties and growth rates of the continuum, and are confined within thin conductive layers around the position of the continuum singularities. The width of the conductive layer, and so the spatial scale of the quasi-continuum modes, depends on the value of the perpendicular thermal conductivity. \citet{vanderlinden91} and \citet{vanderlinden93} applied these results to the context of solar prominences and showed that the spatial scales related to the most unstable quasi-continuum modes of the spectrum are consistent with the size of the prominence threads reported from high-resolution observations. This suggests that perpendicular thermal conduction may be responsible for the formation of prominence fine structure. In subsequent works, \citet{ireland92,ireland98} investigated the influence of magnetic diffusion on the thermal continuum. These authors concluded that magnetic diffusion plays a similar role to that of perpendicular thermal conduction as both mechanisms replace the continuum by discrete modes. 
 
In the present work we follow the method of \citet{vanderlindencont} and \citet{vanderlinden91} and investigate the thermal stability of an inhomogeneous prominence thread model. The equilibrium model adopted here is a cylindrical magnetic tube, representing a prominence thread, embedded in a coronal environment. The plasma physical properties, i.e., temperature, density, etc., are inhomogeneous in the transverse direction, and vary from prominence to PCTR and coronal values. In this investigation, we assume that the equilibrium is uniform in the longitudinal direction. The effect of longitudinal plasma inhomogeneity is relegated to a forthcoming study. We use the nonadiabatic and resistive MHD equations and superimpose linear perturbations on the equilibrium state. Parallel and perpendicular thermal conduction, radiative losses, heating, and magnetic diffusion are the nonideal effects included in our equations. We combine both analytical methods and numerical computations to study the properties of the discrete thermal continuum modes, focusing on the unstable modes with largest growth rates.

This paper is organized as follows. Section~\ref{sec:basic} contains the basic equations and a description of the equilibrium configuration. Then, Sections~\ref{sec:therm} -- \ref{sec:param} contain a theoretical study of the properties of the thermal instability. The stability of the thermal continuum in the absence of perpendicular thermal conduction and magnetic diffusion is discussed in Section~\ref{sec:therm} by following the analysis of \citet{vanderlindencont}, while the unstable thermal modes are investigated in Section~\ref{sec:thermalmodes} when both perpendicular thermal conduction and magnetic diffusion are included. A parametric study of the growth rate of the most unstable solution is performed in Section~\ref{sec:param}. Then, we discuss in Section~\ref{sec:discussion} the physical implication of our theoretical results for the stability and lifetime of prominence threads. Finally, the summary of our results is given in Section~\ref{sec:sum}.

\section{BASIC EQUATIONS AND EQUILIBRIUM}
\label{sec:basic}

%

The basic MHD equations governing the dynamics of a nonadiabatic and resistive plasma are
\begin{equation}
  \frac{{\rm D} \rho}{{\rm D} t} + \rho \nabla \cdot {\bf v} = 0, \label{eq:continuitysum}
\end{equation}
\begin{equation}
   \rho \frac{{\rm D} \bf v}{{\rm D} t} = - \nabla p +  \frac{1}{\mu} \left( \nabla \times {\bf B} \right) \times { \bf B},   \label{eq:motionsum}
\end{equation}
\begin{eqnarray}
 \frac{\partial {\bf B}}{\partial t} = \nabla \times \left( {\bf v} \times {\bf B} \right) - \nabla \times \left(\eta \nabla \times  {\bf B} \right), \label{eq:inductionsum}
\end{eqnarray}
 \begin{equation}
   \frac{{\rm D} p}{{\rm D} t} - \frac{\gamma p}{\rho} \frac{{\rm D}\rho}{{\rm D} t} + \left( \gamma -1 \right)\left[\rho L(T,\rho) -\nabla \cdot \left( {\bf \kappa} \cdot \nabla T \right)   -   \eta | \nabla \times {\bf B}|^2  \right] = 0, \label{eq:energysum}
 \end{equation}
\begin{equation}
 p = \rho R \frac{T}{\mutilde}, \label{eq:statesum}
\end{equation}
along with the condition $\nabla \cdot {\bf B} = 0$, where $\frac{\rm D}{{\rm D} t} = \frac{\partial}{\partial t} + {\bf v} \cdot \nabla$ is the material derivative for time variations following the motion. In Equations~(\ref{eq:continuitysum})--(\ref{eq:statesum}), $\rho$, $p$, $T$, $\bf v$, and $\bf B$ are the mass density, gas pressure, temperature, velocity vector, and magnetic field vector, respectively. In addition, $\mu = 4\pi \times 10^{-7}$~N~A$^{-2}$ is the magnetic permeability, $\gamma = 5/3$ is the adiabatic index, $R=8.3 \times 10^3$~m$^2$~s$^{-2}$~K$^{-1}$ is the ideal gas constant, $L(T,\rho) $ is the heat-loss function, ${\bf \kappa} $ is the thermal conductivity tensor, $\eta$ is the magnetic diffusivity, and $\mutilde$ is the mean atomic weight. In a fully ionized medium, $\mutilde =0.5$.

\subsection{Nonideal terms}

The induction (Equation~(\ref{eq:inductionsum})) and energy (Equation~(\ref{eq:energysum})) equations contain several nonideal terms whose physical meaning is explained next. The term with the factor $\eta$ in Equation~(\ref{eq:inductionsum}) corresponds to Ohm's magnetic diffusion, while the equivalent term in Equation~(\ref{eq:energysum}) accounts for Ohm's heating. The other nonideal terms in Equation~(\ref{eq:energysum}) correspond to nonadiabatic effects, i.e., thermal conduction, radiative losses, and an arbitrary heating input. 

As for thermal conduction, the thermal conductivity tensor, $\bf \kappa$, can be expressed in terms of its parallel, $\kappa_\parallel$, and perpendicular, $\kappa_\perp$, components to the magnetic field, namely ${\bf \kappa} = \kappa_\parallel \hat{e}_B \hat{e}_B + \kappa_\perp \left( \hat{\bf I} - \hat{e}_B \hat{e}_B \right)$, where $\hat{e}_B = {\bf B}/\left| {\bf B} \right|$ is the unit vector in the magnetic field direction and $\hat{\bf I}$ is the identity tensor. The parallel and perpendicular conductivities are dominated by the effect of electrons and ions, respectively. The perpendicular electron conductivity and the parallel ion conductivity are negligible in fully ionized plasmas. Expressions for the conductivities in MKS units for a hydrogen plasma are \citep[see, e.g.,][]{parker,spitzer,brag},
\begin{eqnarray}
 \kappa_\parallel &=& 1.8 \times 10^{-10} \frac{ T^{5/2}}{\ln \Lambda}, \\
 \kappa_\perp&=& 1.48 \times 10^{-42} \frac{\ln\Lambda\,  \rho^2}{m_{\rm i}^2 |{\bf B}|^2 T^{1/2}}, \label{eq:ionskappa}
\end{eqnarray}
 where $m_{\rm i}$ is the proton mass and $\ln \Lambda$ is the Coulomb logarithm, whose value is generally between 5 and 20 and has a weak dependence on temperature and density \citep[see, e.g.,][]{priest}.

Finally, the heat-loss function  $L(T,\rho)$ accounts for the balance between radiative cooling and heating. The determination of an analytical function of the temperature and density that describes radiative losses of the prominence plasma is a very difficult work that requires the numerical solution of nonlocal thermodynamic equilibrium (NLTE) radiative transfer equations. This is beyond the purpose and scope of the present study. One reasonable semi-empirical approximation to an expression for the radiative loss function was obtained by \citet{hildner}. This author assumed an optically thin plasma \citep[e.g.,][]{coxtucker} and performed a piecewise fit of the radiative losses previously computed by several authors as a function of temperature. Here, we adopt Hildner's approach.  The functional expression of $L(T,\rho)$ considered by \citet{hildner} is
\begin{equation}
 L(T,\rho) = \rho \chi^*  T^\alpha - h, \label{eq:radlosses}
\end{equation}
where $\chi^*$ and $\alpha$ are piecewise constants depending on the temperature, and $h$ is an arbitrary heating function. The assumption of an optically thin plasma seems a reasonable approximation for coronal and prominence-corona transition region (PCTR) temperatures, whereas cool prominence plasmas may be considered optically thick. Some authors \citep[e.g.,][]{rosner,milne} have proposed corrections to the values of $\chi^*$ and $\alpha$ in the range of cool prominence temperatures, i.e., $T < 15,000$~K, in order to represent radiation losses in optically thick plasmas using Equation~(\ref{eq:radlosses}). In the literature, there are more recent parametrizations for $\chi^*$ and $\alpha$ which update Hildner's values. In particular, the so-called Klimchuk-Raymond parametrization \citep[see, e.g.,][]{klimchuk} may be a better representation of the radiative losses in the hotter part of the PCTR and in the solar corona. On the contrary, the Klimchuk-Raymond function may not be adequate in the cool part of the thread. We use the Klimchuk-Raymond function as an alternative to Hildner's parametrization in the hotter part of the equilibrium. The values  of the parameters $\chi^*$ and $\alpha$ for various temperature ranges and regimes are given in Table~\ref{tab:regimes}.

\begin{table}[!t]   
\centering                         
\begin{tabular}{l c c r }      
\hline\hline                 
Regime & Temperature range &$\chi^*$ & $\alpha$  \\  
\hline                      
   Prominence-1.1 & $T \leq$ 15$\times 10^3$~K & $1.76 \times 10^{-13}$ & $7.4$   \\     
   Prominence-1.2 &  $T \leq$ 15$\times 10^3$~K & $1.76 \times 10^{-53}$ &  $17.4$   \\
   Prominence-1.3 &  $T \leq$ 15$\times 10^3$~K & $7.01 \times 10^{-104}$ &  $30$    \\
   PCTR-2 &  15$\times 10^3$~K $<T \leq$ 8$\times 10^4$~K  & $4.29 \times 10^{10}$ & $1.8$  \\
   PCTR-3 & 8$\times 10^4$~K $<T \leq$ 3$\times 10^5$~K & $2.86 \times 10^{19}$ & $0.0$  \\
   PCTR-4 & 3$\times 10^5$~K $<T \leq$ 8$\times 10^5$~K & $1.41 \times 10^{33}$ & $-2.5$  \\
   Corona-5  & $T>$ 8$\times 10^5$~K  & $1.97 \times 10^{24}$ & $-1.0$  \\ 
\hline
Klimchuk-Raymond-1 & $T \leq 10^{4.97}$~K & $3.91 \times 10^{9}$  & $2.0$ \\
Klimchuk-Raymond-2 & $10^{4.97} < T \leq 10^{5.67}$~K & $3.18 \times 10^{24}$ & $-1.0$ \\
Klimchuk-Raymond-3 & $ T > 10^{5.67}$~K & $6.81 \times 10^{18}$ & $0.0$ \\
\hline                                   
\end{tabular}
\caption{Values in MKS units of the parameters in the radiative loss function (Equation~(\ref{eq:radlosses})) corresponding to several temperature regimes. Prominence-1.1, PCTR-1, PCTR-2, and PCTR-3, and Corona-5 regimes are parametrizations from \citet{hildner}. Prominence-1.2 and Prominence-1.3 regimes are taken from \citet{milne} and \citet{rosner}, respectively. The three Prominence regimes represent different plasma optical thicknesses, Prominence-1.1 corresponding to optically thin plasma, while Prominence-1.2 and Prominence-1.3 are for optically thick and very thick plasmas, respectively. The three Klimchuk-Raymond regimes are adapted from \citet{klimchuk}, where we only have taken into account the range of temperatures considered in our equilibrium.  \label{tab:regimes} }             
\end{table}

\subsection{Equilibrium configuration}

Equations~(\ref{eq:continuitysum})--(\ref{eq:statesum}) are applied to the following equilibrium configuration. The prominence thread model is composed of a straight and cylindrical magnetic flux tube embedded in a coronal environment. We use cylindrical coordinates, namely $r$, $\varphi$, and $z$ for the radial, azimuthal, and longitudinal coordinates, respectively. In this configuration, the equilibrium quantities are invariant in the azimuthal and longitudinal directions, so they depend on the radial direction only. Hereafter, the equilibrium quantities are denoted by a subscript 0. The equilibrium magnetic field is straight and homogeneous, ${\bf B}_0 = B_0 \hat{e}_z$, where, for simplicity, $B_0$ is the same constant everywhere. Magnetic twist and longitudinal plasma inhomogeneity are effects that will be included in future investigations.

Since the magnetic field is straight and homogeneous, the equilibrium gas pressure, $p_0$, is also homogeneous. The equilibrium density, $\rho_0$, and temperature, $T_0$, profiles must verify the condition of energy balance, which from Equation~(\ref{eq:energysum}) is
\begin{equation}
 \frac{1}{r} \frac{\rm d}{{\rm d} r} \left( r  \kappa_\perp \frac{{\rm d} T_0}{{\rm d} r} \right) = \rho_0^2 \chi^*  T^\alpha_0 - \rho_0 h , \label{eq:balance}
\end{equation}
where we have assumed that there are no flows in the equilibrium, i.e., ${\bf v}_0 = {\bf 0}$. Note that for the present equilibrium magnetic field, Ohm's heating term is absent from Equation~(\ref{eq:balance}). The density is related to the temperature and gas pressure through Equation~(\ref{eq:statesum}). We can use Equation~(\ref{eq:balance}) to obtain the temperature profile and later use Equation~(\ref{eq:statesum}) to compute the density profile. Unfortunately, the heating function $h$ is unknown in prominences and in the corona, which makes it impossible to apply this procedure. Instead, we follow an alternative method that has been adopted in a number of previous works \citep[e.g.,][]{vanderlinden91,ireland92,ireland98}. We choose an ad-hoc temperature profile and compute from Equation~(\ref{eq:balance}) the corresponding heating function that satisfies energy balance. The temperature profile adopted in this work is in agreement with theoretical models of prominence threads  \citep[see, e.g.,][Fig.~12]{cirigliano}.

Thus, the equilibrium temperature profile is
\begin{equation}
 T_0 \left( r \right) = \left\{
\begin{array}{lll}
 T_{\rm p}, & \textrm{if} & r \leq R - l /2, \\
T_{\rm PCTR} \left( r \right), & \textrm{if} & R - l /2 < r < R + l /2,\\
T_{\rm c}, & \textrm{if} & r \geq R + l /2.
\end{array}
\right. 	\label{eq:profilegen}
\end{equation}
We consider that the prominence thread is composed by a central and cool region with homogeneous temperature, $T_{\rm p}$, surrounded by a transverse transitional layer where the temperature abruptly increases until the homogeneous coronal medium with temperature $T_{\rm c}$ is reached. Following the idea explained by \citet{cirigliano}, this transitional layer represents the particular PCTR of the prominence thread. The mean radius of the thread is denoted by $R$ and is located at the center of the PCTR. The thickness of the PCTR is denoted by $l$, and corresponds to the region $R - l /2 < r < R + l /2$. So, the cases $l/R = 0$ and $l/R = 2$ correspond to a thread without transitional layer and a radially full inhomogeneous thread, respectively. In Equation~(\ref{eq:profilegen}) and in the following expressions, the subscripts p, PCTR, and c denote the central prominence region, the transitional layer, and the coronal medium, respectively. Unless otherwise stated, we consider the following values of the equilibrium quantities: $T_{\rm p} = 8000$~K, $T_{\rm c} = 10^6$~K, $B_0 = 10$~G, and $R=100$~km. The central prominence density is fixed to $\rho_{\rm p} = 5 \times 10^{-11}$~kg~m$^{-3}$, thus the gas pressure is $p_0 = 6.64\times 10^{-3}$~Pa, which corresponds to $\beta = p_0 / \left( B_0^2 / 2 \mu \right) = 0.008$.

\begin{figure}[!t]
\centering
 \epsscale{0.49}
 \plotone{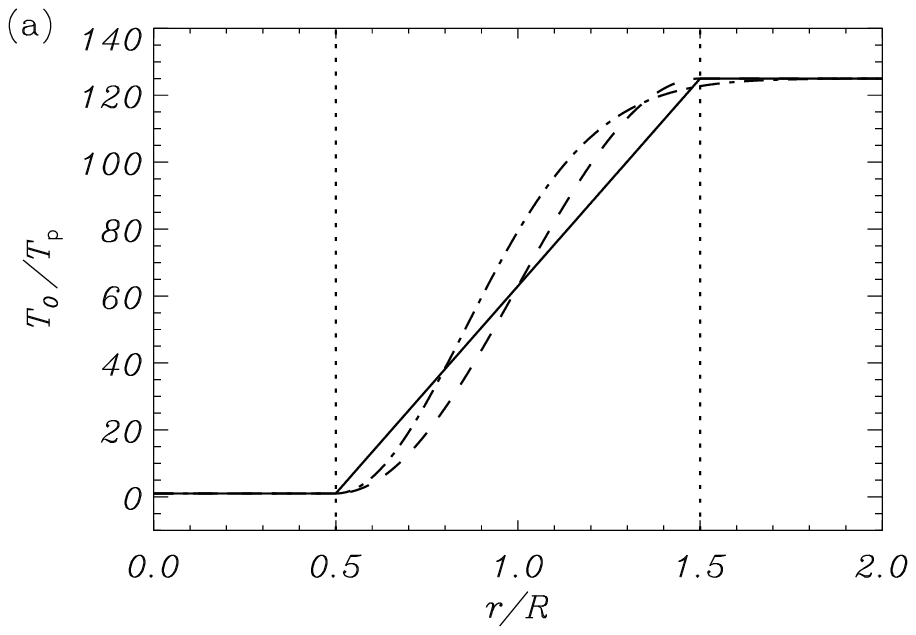}
 \plotone{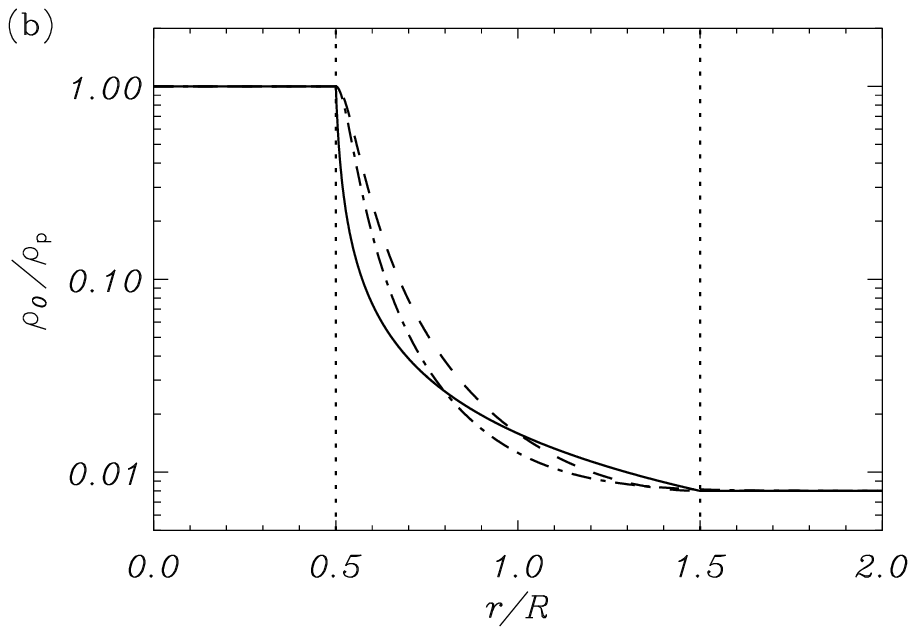}
\caption{(a) Equilibrium temperature profiles, $T_0$, normalized to the internal temperature, $T_{\rm p}$, in the case $l/R = 1$. The different line styles correspond to the linear profile (solid), the sinusoidal profile (dashed), and the Gaussian profile (dot-dashed). (b) Corresponding density profiles, $\rho_0$, normalized to the internal density, $\rho_{\rm p}$. The vertical dotted lines in both panels denote the boundaries of the transitional zone. Note that the vertical axis in panel (b) is in logarithmic scale. \label{fig:profile}}
\end{figure}

In order to assess whether or not the form of the temperature profile influences the results, we consider three different expressions of $T_{\rm PCTR} \left( r \right)$, which correspond to a linear profile,
\begin{equation}
 T_{\rm PCTR} \left( r \right) = T_{\rm p} + \frac{T_{\rm c} - T_{\rm p}}{l} \left( r - R + l/2 \right),
\end{equation}
a sinusoidal profile,
\begin{equation}
 T_{\rm PCTR} \left( r \right) = \frac{T_{\rm p}}{2} \left\{ \left( 1 + \frac{T_{\rm c}}{T_{\rm p}} \right) - \left( 1 - \frac{T_{\rm c}}{T_{\rm p}}  \right) \sin \left[ \frac{\pi}{l} \left( r - R \right)  \right]  \right\},
\end{equation}
and a Gaussian profile
\begin{equation}
  T_{\rm PCTR} \left( r \right) = T_{\rm c} - \left( T_{\rm c} - T_{\rm p} \right) \exp \left[ - \left( \frac{ r  - R + l/2}{l/2} \right)^2\right]. \label{eq:gaussian}
\end{equation}
In the case of the Gaussian profile (Equation~(\ref{eq:gaussian})), note that the exponential factor is $\exp \left( -4 \right) \approx 0.018$ for $r = R + l /2$, so the temperature is not strictly $T_{\rm c}$ when the corona is reached. For consistency, Equation~(\ref{eq:gaussian}) also applies for $r > R + l /2$ in order to avoid this small jump of the temperature when the Gaussian profile is used. Figure~\ref{fig:profile} displays a plot of the three different temperature profiles assumed in this work and their corresponding density profiles.

\subsection{Linear perturbations}

We superimpose perturbations on the equilibrium state. Hence, $\rho_1$, $T_1$, and $p_1$ denote the density, temperature, and gas pressure perturbations, respectively, whereas ${\bf v}_1 = \left( v_r, v_\varphi, v_z \right)$ and ${\bf b}_1 = \left( b_r, b_\varphi, b_z \right)$ are the velocity and magnetic field perturbations. Next, we assume that these perturbations are small, so we restrict ourselves to the linear regime and Equations~(\ref{eq:continuitysum})--(\ref{eq:statesum}) are linearized. Since the equilibrium is invariant in the azimuthal and longitudinal directions, we write all perturbations proportional to $\exp \left( s t + i m \varphi - i k_z z \right)$, where $m$ and $k_z$ are the azimuthal and longitudinal wavenumbers, respectively, and $s$ is the growth (or damping) rate of the perturbation. Equations~(\ref{eq:continuitysum})--(\ref{eq:statesum}) become,
\begin{equation}
 s \rho_1 = -\rho_0' v_r - \rho_0 \left( v_r' + \frac{1}{r} v_r + \frac{m}{r} v_\varphi + k_z v_z \right), \label{eq:lin1}
\end{equation}
\begin{equation}
 s v_r = - \frac{\cs^2}{\gamma} \left( \frac{\rho_1'}{\rho_0} + \frac{T_1'}{T_0} - \frac{\rho_0'}{\rho_0^2} \rho_1 - \frac{T_0'}{T_0^2} T_1 \right) + \frac{\va^2}{B_0} \left( k_z b_r - b_z' \right), 
\end{equation}
\begin{equation}
 s v_\varphi =  \frac{\cs^2}{\gamma} \frac{m}{r} \left( \frac{\rho_1}{\rho_0} + \frac{T_1}{T_0} \right) - \frac{\va^2}{B_0} \left( k_z b_\varphi - \frac{m}{r} b_z \right),
\end{equation}
\begin{equation}
 s v_z = \frac{\cs^2}{\gamma} k_z \left( \frac{\rho_1}{\rho_0} + \frac{T_1}{T_0} \right),
\end{equation}
\begin{equation}
 s b_r = - B_0 k_z v_r + \eta \left( \frac{m}{r}  b_\varphi' + \frac{m}{r^2} b_\varphi - \frac{m^2}{r^2} b_r - k_z^2 b_r + k_z b_z'  \right),
\end{equation}
\begin{eqnarray}
 s b_\varphi =&& B_0 k_z v_\varphi \nonumber \\ &+& \eta \left( b_\varphi'' + \frac{1}{r} b_\varphi' - \frac{1}{r^2} b_\varphi -  \frac{m}{r} b_r' +   \frac{m}{r^2} b_r - k_z^2 b_\varphi + k_z  \frac{m}{r} b_z \right) \nonumber \\ 
&-& \eta' \left( \frac{m}{r} b_r - b_\varphi' - \frac{1}{r} b_\varphi  \right),
\end{eqnarray}
\begin{eqnarray}
 s b_z = &-& B_0 \left( v_r' + \frac{1}{r} v_r + \frac{m}{r} v_\varphi \right) \nonumber \\ &+& \eta \left( b_z'' + \frac{1}{r} b_z' - \frac{m^2}{r^2} b_z + \frac{m}{r} k_z b_\varphi  - \frac{k_z}{r} b_r - k_z b_r' \right) \nonumber \\ &+& \eta' \left( b_z' - k_z b_r  \right),
\end{eqnarray}
\begin{eqnarray}
 &s& \left[ \frac{p_0}{T_0} T_1 - \left( \gamma - 1 \right) \frac{p_0}{\rho_0} \rho_1 \right] = \cs^2 \rho_0' v_r \nonumber \\
 &-& \left( \gamma - 1 \right) \left[ \frac{p_0}{\rho_0} \omega_\rho \rho_1 + \left(  \frac{p_0}{T_0} \omega_T + \kappa_\perp \frac{m^2}{r^2} + \kappa_\parallel k_z^2 \right) T_1 \right] \nonumber \\
 &+& \left( \gamma - 1 \right) \left[ \left( \kappa_\perp \frac{1}{r} + \kappa_\perp'  \right) T_1' +  \kappa_\perp T_1''  + \left(  \kappa_\parallel - \kappa_\perp \right) \frac{T_0'}{B_0} k_z b_r \right]\nonumber \\
&+& \left( \gamma - 1 \right) \left[  \left( T_0'' + \frac{1}{r} T_0 \right) \tilde{\kappa}_{\perp}  + T_0' \tilde{\kappa}_{\perp}' \right],\label{eq:linfin}
\end{eqnarray}
where the prime denotes the derivative with respect to $r$, $\cs^2 = \frac{\gamma p_0}{\rho_0}$ is the sound speed squared, $\va^2 = \frac{B_0}{\mu \rho_0}$ is the Alfv\'en speed squared, and the quantities $\omega_\rho$, $\omega_T$, and $\tilde{\kappa}_{\perp}$ are defined as follows,
\begin{equation}
 \omega_\rho = \frac{\rho_0}{p_0} \left[ L\left( \rho_0, T_0 \right) + \rho_0 \left( \frac{\pd L}{\pd \rho} \right)_{\rho_0,T_0} \right],  \label{eq:wrho}
\end{equation}
\begin{equation}
 \omega_T = \frac{\rho_0}{p_0} T_0 \left( \frac{\pd L}{\pd T} \right)_{\rho_0,T_0},  \label{eq:wt}
\end{equation}
\begin{equation}
 \tilde{\kappa}_{\perp} = \left( \frac{\pd \kappa_\perp}{\pd \rho} \right)_{\rho_0,T_0} \rho_1 + \left( \frac{\pd \kappa_\perp}{\pd T} \right)_{\rho_0,T_0} T_1 + \left( \frac{\pd \kappa_\perp}{\pd \left| {\bf B_0} \right|} \right)_{\rho_0,T_0} \frac{{\bf B_0} \cdot {\bf b_1}}{\left| {\bf B_0} \right|}.
\end{equation}

Equations~(\ref{eq:lin1})--(\ref{eq:linfin}) form an eigenvalue problem, with $s$ the eigenvalue. To eliminate one variable and simplify matters, we have used the linearized version of Equation~(\ref{eq:statesum}) to write the gas pressure perturbation, $p_1$, in terms of the density, $\rho_1$, and the temperature, $T_1$, perturbations. In addition, to express Equations~(\ref{eq:lin1})--(\ref{eq:linfin}) in terms of real quantities, we have performed the substitutions $i v_\varphi \to v_\varphi$, $i v_z \to v_z$, and  $i b_r \to b_r$. Note that the term related to Ohm's heating in Equation~(\ref{eq:energysum}) does not contribute in the linear regime and so it is absent from the linearized energy equation (Equation~(\ref{eq:linfin})). In addition, since the heating function $h$ represents an external, arbitrary heating source, no perturbations of this function have been considered.

We numerically solve Equations~(\ref{eq:lin1})--(\ref{eq:linfin}) and compute the eigenvalues and their corresponding perturbations by means of the PDE2D code \citep{sewell} based on finite elements.  The numerical integration of Equations~(\ref{eq:lin1})--(\ref{eq:linfin}) is performed from the thread axis, $r=0$, to the finite edge of the numerical domain, $r=r_{\rm max}$, which is located far enough to obtain a good convergence of the solution and to avoid numerical errors. We use a nonuniform grid with a large density of grid points within the inhomogeneous transitional layer, where the equilibrium properties vary by several orders of magnitude. The nonuniform grid also allows us to correctly describe the small spatial scales of the eigenfunctions within the transitional layer. The PDE2D code uses a collocation method and the generalized matrix eigenvalue problem is solved using the shifted inverse power method. The output of the program is the closest eigenvalue to an initial provided guess and its corresponding perturbations.

In the present investigation, we restrict ourselves to the study of thermal quasi-continuum eigenmodes. For real $k_z$ and $m$, these solutions correspond to real values of $s$. When the eigenmode is a stable, damped solution, $s < 0$, while unstable thermal modes have $s > 0$. As shown by \citet{vanderlinden91} and \citet{ireland92}, some of the basic properties regarding the stability of the thermal quasi-continuum modes can be deduced from the properties of the thermal continuum in the absence of perpendicular thermal conduction and magnetic diffusion. We perform this analysis in the next Section.

\section{THE THERMAL CONTINUUM}

\label{sec:therm}

In our equilibrium configuration, the full spectrum of ideal eigenmodes, i.e., when nonideal effects are absent, contains solutions related to two different continua, namely the Alfv\'en and slow (or cusp) continua \citep[see, e.g.,][]{appert}. These continua are modified or removed if nonideal effects are taken into account. The effect of the different nonideal mechanisms on the continua has been studied by a number of authors, whose relevant results for our investigation are briefly summarized next.

If magnetic diffusion is included but nonadiabatic terms are neglected, both Alfv\'en and slow continua are replaced by a set of discrete modes with complex $s$ and whose properties depend on the value of the diffusivity \citep[see, e.g.,][]{stefaan}. In the presence of radiation losses and parallel thermal conduction but for $\kappa_\perp = \eta = 0$, the Alfv\'en continuum remains unaffected and the slow continuum is only slightly modified becoming a nonadiabatic slow continuum with complex $s$. In addition, a new thermal continuum is present, which corresponds to real values of $s$ \citep{vanderlindencont}. When perpendicular thermal conduction is included, the thermal continuum is replaced by a dense set of eigenmodes, i.e., a quasi-continuum \citep{vanderlinden91}. Finally, if both perpendicular thermal conduction and magnetic diffusion are taken into account, the three continua are removed and replaced by discrete eigenmodes \citep{ireland92}. 

To study the properties of the thermal continuum before solving the full problem numerically, we fix $\kappa_\perp = \eta  = 0$ in Equations~(\ref{eq:lin1})--(\ref{eq:linfin}). After a lengthy but straightforward process, it is possible to combine Equations~(\ref{eq:lin1})--(\ref{eq:linfin}) to obtain the two coupled, first-order differential equations of \citet[Equations~(2)--(3)]{vanderlindencont}, which depend on the coefficient $C_0$ defined as
\begin{equation}
C_0 = r \frac{\rho_0^3 p_0}{T_0}  \left( s^2 + k_z^2 \va^2 \right) C_t, \label{eq:c0}
\end{equation}
with $C_t$ the following third-order polynomial in $s$,
\begin{eqnarray}
 C_t &=& \frac{\cs^2 + \va^2}{\gamma - 1} s^3 + \left[ \left( \frac{\cs^2}{\gamma} + \va^2  \right) \left(  \frac{T_0}{p_0} \kappa_\parallel k_z^2 + \omega_T \right) - \frac{p_0}{\rho_0} \omega_\rho \right] s^2 \nonumber \\ &+& \frac{\cs^2 \va^2}{\gamma - 1} k_z^2 s 
+ \frac{\cs^2}{\gamma} \va^2 k_z^2 \left( \frac{T_0}{p_0} \kappa_\parallel k_z^2 + \omega_T - \omega_\rho \right). \label{eq:ct}
\end{eqnarray}
\citet{vanderlindencont} showed that the roots of $C_0 = 0$, i.e., the singularities in their Equations~(2)--(3), correspond to the three continua.  Note that Equations~(\ref{eq:c0}) and (\ref{eq:ct}) are independent of the azimuthal wavenumber $m$, and so the three continua do not depend on the value of $m$.  For a fixed $r$, the roots of $C_0 = 0$ are two purely imaginary solutions given by $s^2 = - k_z^2 \va^2$, that correspond to the Alfv\'en continuum, and the solutions of $C_t = 0$. Since $C_t$ is a third-order polynomial and the nonadiabatic terms are assumed to be small, the roots of $C_t = 0$ are two complex conjugate solutions corresponding to the nonadiabatic slow continuum, and a real root corresponding to the thermal continuum. It is straightforward to check that in the ideal case $\kappa_\parallel = \omega_\rho = \omega_T = 0$, so $C_t$ becomes
\begin{equation}
C_t = \frac{s}{\gamma -1} \left[ \left( \cs^2 + \va^2  \right) s^2 + \cs^2 \va^2 k_z^2 \right]. \label{eq:ctideal}
\end{equation}
The roots of $C_t = 0$ are then $s^2 = - \frac{\cs^2 \va^2}{\cs^2 + \va^2} k_z^2$, which correspond to the ideal slow continuum, while the thermal continuum disappears and becomes the trivial solution $s=0$.

\subsection{Approximate growth rates and stability criterion}

\begin{figure}[!t]
\centering
 \epsscale{0.49}
 \plotone{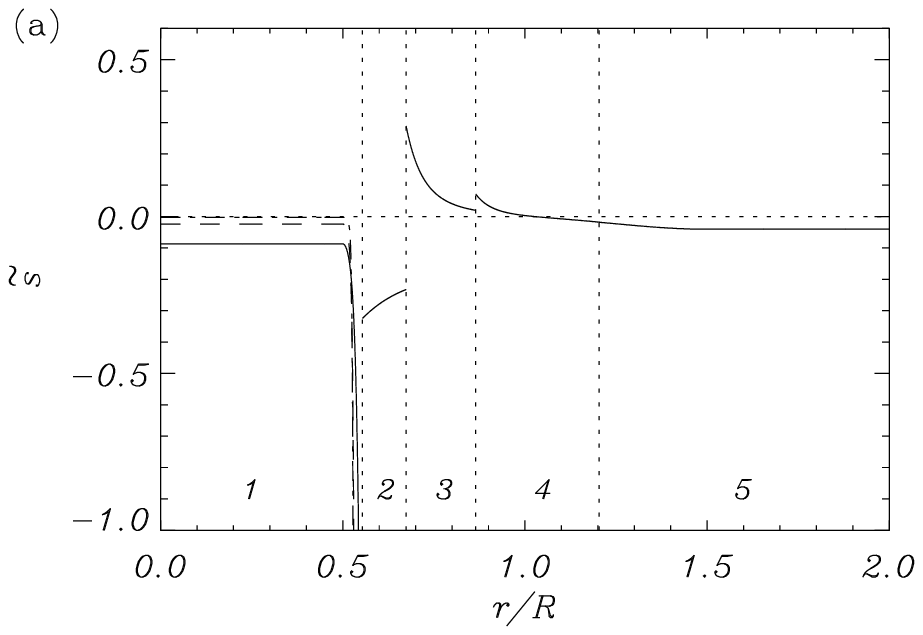}
 \plotone{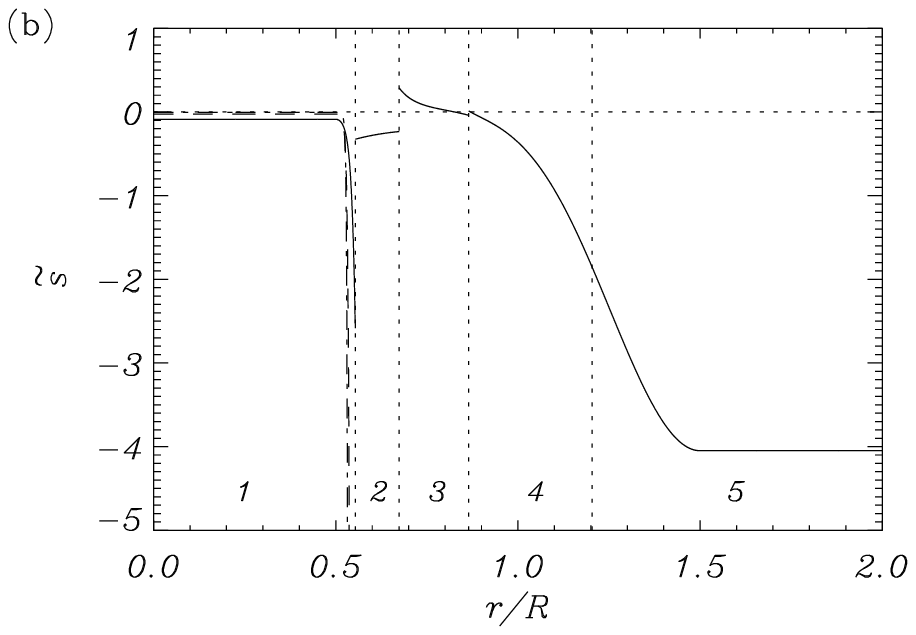}
\caption{Normalized thermal continuum as a function of $r/R$ in the case of the sinusoidal temperature profile with $l/R=1$ for (a) $k_z R = 10^{-2}$ and (b) $k_z R = 10^{-1}$. The vertical dotted lines denote the boundaries between the different parametrizations of the radiation function. Each zone is indicated by a number from 1 to 5, that corresponds to the different regimes of Hildner's piecewise fit (Table~\ref{tab:regimes}). The horizontal dotted line represents $\tilde{s}=0$. The different line styles in zone 1 correspond to the Prominence-1.1 (solid), Prominence-1.2 (dashed), and Prominence-1.3 (dot-dashed) parametrizations.  \label{fig:continuum}}
\end{figure}

An approximation to the thermal continuum growth rate can be obtained by neglecting the terms with $s^2$ and $s^3$ in Equation~(\ref{eq:ct}). Then, the equation $C_t = 0$ gives the approximate solution
\begin{equation}
 s \approx - \frac{\gamma - 1}{\gamma}\left( \frac{T_0}{p_0} \kappa_\parallel k_z^2 + \omega_T - \omega_\rho \right). \label{eq:cont}
\end{equation}
Equation~(\ref{eq:cont}) is equivalent to Equation~(4.12) of \citet{vanderlinden91}. It is worth mentioning that Equation~(\ref{eq:cont}) is similar to the expression obtained by \citet{carbonell09} and \citet{solerphd} for the imaginary part of the frequency of propagating thermal waves in a flowing medium.

The sign of the constant term in Equation~(\ref{eq:ct}) provides us with the instability criterion of the thermal continuum, i.e., the combination of parameters that causes $s > 0$. When we use the definitions of $\omega_\rho$ and $ \omega_T$ (Equations~(\ref{eq:wrho}) and (\ref{eq:wt})) and take into account that the cooling function at equilibrium is $L\left( \rho_0, T_0 \right) = 0$ when $\kappa_\perp = 0$, the instability criterion is
\begin{equation}
 \kappa_\parallel k_z^2 + \rho_0  \left( \frac{\pd L}{\pd T} -  \frac{\rho_0}{T_0} \frac{\pd L}{\pd \rho} \right) < 0, \label{eq:crit2}
\end{equation}
which turns out to be the same instability criterion given by \citet{field} in his Equation~(25a), although Field's criterion was derived for thermal modes in a homogeneous medium and our equilibrium quantities depend on $r$. Therefore, the stability of the continuum also depends on $r$, and Equation~(\ref{eq:crit2}) must be computed for all $r$ to assess the absolute instability of the continuum. In addition, magnetic diffusion may play a role for stability. This issue was addressed by \citet{ireland92}, who showed that, when magnetic diffusion is included, the contribution of Ohm's heating to the equilibrium energy balance has to be taken into account. However, since the magnetic field is homogeneous in our case, Ohm's heating plays no role in the energy balance of our equilibrium (Equation~(\ref{eq:balance})). This fact points out that, in our present application, Equation~(\ref{eq:crit2}) remains valid even when magnetic diffusion is present, and the thermal continuum growth rates are not modified by diffusion.

Equation~(\ref{eq:crit2}) can be rewritten in terms of parameters $\chi^*$ and $\alpha$ of the cooling function, namely 
\begin{equation}
 \kappa_\parallel k_z^2 +\left( \alpha - 1 \right) \rho_0^2  \chi^* T_0^{\alpha-1} < 0.\label{eq:crit1}
\end{equation}
In the absence of parallel thermal conduction, i.e., $\kappa_\parallel = 0$, the instability criterion is satisfied for $\alpha < 1$. According to the different radiative regimes of Table~\ref{tab:regimes}, the condition $\alpha < 1$ takes place for $T_0 >$~ 80,000~K in Hildner's parametrization, meaning that the thermal continuum is unstable in the PCTR-3, PCTR-4, and Corona-5 regimes. Considering the Klimchuk-Raymond fit, instability is present for $T_0 \gtrsim$~93,325~K.  This result agrees with the instabilities found by \cite{vanderlinden91} in the range of coronal temperatures studied by these authors. When parallel thermal conduction is present, the unstable part of the continuum is stabilized by a critical value of the longitudinal wavenumber, namely $k_z^*$, given by
\begin{equation}
 k_z^* = \left[ \frac{1 - \alpha}{\kappa_\parallel}  \rho_0^2  \chi^* T_0^{\alpha-1} \right]^{1/2}. \label{eq:kzcrit}
\end{equation}
Note that $k_z^*$ is also a function of $r$ and Equation~(\ref{eq:kzcrit}) only applies when $\alpha < 1$. The absolute stability of the continuum is guaranteed for $k_z$ larger than the maximum value of $k_z^*$ in the equilibrium.


\begin{figure}[!t]
\centering
 \epsscale{0.49}
 \plotone{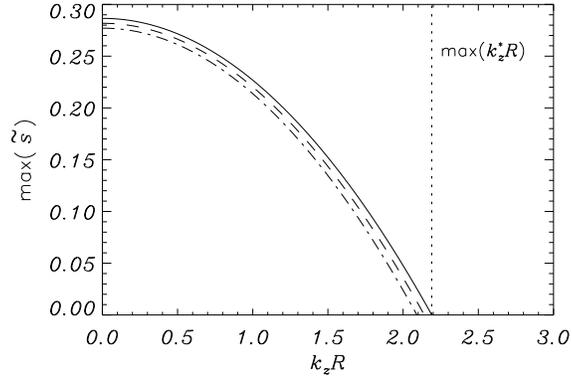}
\caption{Maximum growth rate of the normalized thermal continuum as a function of $k_z R$. The different line styles represent the result for the linear (solid), sinusoidal (dashed), and Gaussian (dot-dashed) temperature profiles with $l/R=1$. The vertical dotted line corresponds to the maximum value of the dimensionless critical wavenumber $k_z^* R$ for the linear temperature profile. Hildner's parametrization for the radiative loss function has been used. \label{fig:continuum2}}
\end{figure}

\begin{figure}[!t]
\centering
 \epsscale{0.49}
 \plotone{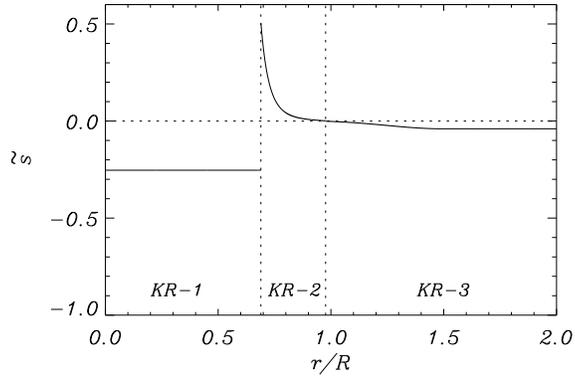}
\caption{Same as Figure~\ref{fig:continuum}(a) but using Klimchuk-Raymond's fit for the radiative loss function. \label{fig:continuumklim}}
\end{figure}

\begin{figure*}[!htb]
\centering
 \epsscale{0.99}
 \plotone{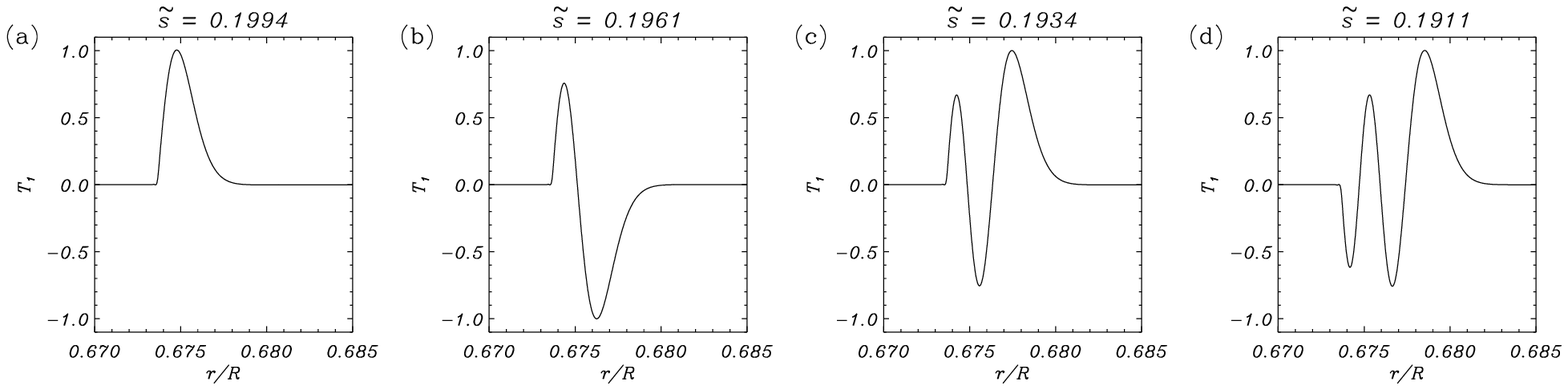}
\caption{Temperature perturbation (in arbitrary units) of the four most unstable modes (from the left to the right) in the absence of magnetic diffusion. Computations performed considering the sinusoidal temperature profile with $l/R=1$, $k_z R = 10^{-1}$, and $m=0$. The corresponding normalized growth rate, $\tilde{s}$, is displayed on top of each panel. \label{fig:eigen1}}
\end{figure*}

\subsection{Properties of the continuum}

Figure~\ref{fig:continuum} shows the thermal continuum computed from Equation~(\ref{eq:cont}) for our equilibrium configuration in the case of the sinusoidal temperature profile with $l/R=1$, and considering Hildner's parametrization for the radiative loss function.  The growth rate has been plotted in dimensionless form computed as $\tilde{s} = s R /\csp$, where $\tilde{s}$ is the dimensionless growth rate and $\csp$ is the sound speed of the central, homogeneous part of the thread. At first sight, we notice the jumps of the growth rate at the boundaries of the zones where different parametrizations of the cooling function are used. According to Hildner's fit (see Table~\ref{tab:regimes}), the cooling function $L \left( \rho , T  \right)$ is continuous at the boundaries between the zones with different radiation regimes, but the derivatives of $L \left( \rho , T  \right)$ with respect to density and temperature are discontinuous. As Equation~(\ref{eq:cont}) depends on these derivatives, the thermal continuum has the apparent form of five separated continuous spectra. If the actual cooling function of prominences is known and is used here instead of the present parametrization, the jumps would be replaced by continuous but very abrupt variations of the growth rate at the boundaries between the different regimes. We must point out that the jumps in the thermal spectrum do not affect the stability of the solutions \citep[see Figs.~1 and 3 of][where this issue is also commented]{vanderlindencont}.  An alternative radiative cooling function computed numerically \citep[e.g.,][]{schure} instead of the present piecewise parametrization might be considered in future applications.

In Figure~\ref{fig:continuum}(a), corresponding to $k_z R = 10^{-2}$, we see that $\tilde{s} > 0$ in zone 3 and part of zone 4, meaning that the continuum is unstable in these regions. This is consistent with the result from the instability criterion (Equation~(\ref{eq:crit1})). For the particular longitudinal wavenumber considered in Figure~\ref{fig:continuum}(a) parallel thermal conduction can suppress the instability in zone 5 and part of zone 4. If the longitudinal wavenumber is increased to $k_z R = 10^{-1}$ (see Fig.~\ref{fig:continuum}(b)), the instability is also suppressed in zone 4 and part of zone 3. To completely suppress the instability in the whole continuum, we must increase the longitudinal wavenumber until the maximum value of $k_z^* R$, given by Equation~(\ref{eq:kzcrit}) is reached. Figure~\ref{fig:continuum2} displays the maximum growth rate of the spectrum as a function of $k_z R$. As expected, the maximum growth rate decreases as $k_z R$ increases, until absolute stability of the thermal continuum is achieved for a critical longitudinal wavenumber. This critical wavenumber is in perfect agreement with Equation~(\ref{eq:kzcrit}). The result for the three different temperature profiles does not show significant differences. We must mention that the various parametrizations of the radiative regime Prominence-1, which aim to represent different  optical thicknesses of the cool prominence plasma, are not relevant for the instability of the continuum. The thermal spectrum in the cool part of the equilibrium, namely zone 1, is always stable independently of the considered parametrization (indicated by different line styles in Fig.~\ref{fig:continuum})

In Figure~\ref{fig:continuumklim} we have plotted the thermal continuum growth rate for $k_z R = 10^{-2}$ using the Klimchuk-Raymond radiative loss function. By comparing Figures~\ref{fig:continuum}(a) and \ref{fig:continuumklim}, we see that slightly larger values of the growth rate are obtained using Klimchuk-Raymond's fit with respect to the values for Hildner's fit. However, the qualitative behavior of the continuum is similar is both cases.

\section{PROPERTIES OF THE UNSTABLE THERMAL MODES}

\label{sec:thermalmodes}


%

\subsection{Effect of perpendicular thermal conduction}

\label{sec:kappa}

Here, we study how the unstable part of the thermal continuum is modified when perpendicular thermal conduction and magnetic diffusion are taken into account. Unless otherwise stated, Hildner's parametrization for the radiative loss function is used in all the following computations. First, we set $\eta=0$ and focus our investigation on the effect of perpendicular thermal conduction. We would like to stress that the real physical value of $\kappa_\perp$ given by Equation~(\ref{eq:ionskappa}) is used in the following computations. As stated by \citet{vanderlinden91}, the thermal continuum is replaced by a set of discrete modes when $\kappa_\perp \ne 0$. The eigenfunctions of these solutions display large variations in a region surrounding the position of the thermal continuum singularity for  $\kappa_\perp = 0$. Figure~\ref{fig:eigen1} displays the temperature perturbation, $T_1$, of the four most unstable  modes of our equilibrium  in the case of the sinusoidal temperature profile with $l/R = 1$, $k_z R = 10^{-1}$ and $m=0$. For simplicity, Figure~\ref{fig:eigen1} shows the temperature perturbation only, because $T_1$ is between one and two orders of magnitude larger than the other perturbations. This means that the temperature perturbation is the dominant disturbance related to the thermal modes, although these solutions produce also velocity and magnetic field perturbations. Figure~\ref{fig:eigen1} focuses on the region where the eigenfunctions show significant variations, i.e., the conductive layer described by \citet{vanderlinden93}, whereas their amplitude outside the range plotted in Figure~\ref{fig:eigen1} is negligible. When  Figure~\ref{fig:eigen1} is repeated for the linear and Gaussian temperature profiles, we find that the eigenfunctions are shifted toward smaller $r/R$ for the linear profile, and to larger $r/R$ for the Gaussian profile. However, the form of the perturbations is very similar to those displayed in Figure~\ref{fig:eigen1} and, for simplicity, we do not plot the perturbations again.

\begin{figure}[!htp]
\centering
 \epsscale{0.49}
 \plotone{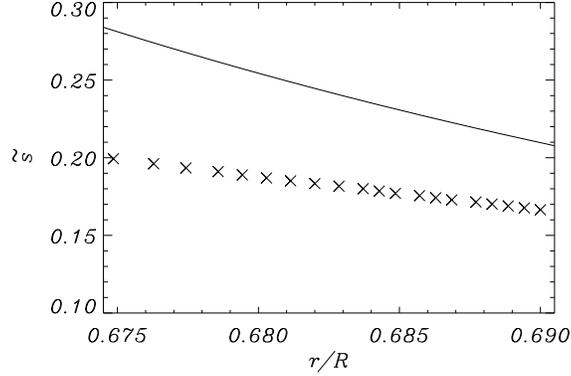}
\caption{Normalized growth rate of the 20 most unstable modes (symbol $\times$) as a function of the position of their largest maximum or minimum. The solid line represents the thermal continuum in the case $\kappa_\perp = 0$. Results are computed for $\eta = 0$, $k_z R = 10^{-1}$, and $m=0$. The sinusoidal temperature profile with $l/R=1$ has been used.  \label{fig:growth20}}
\end{figure}

The order of the solutions can be easily identified by the number of extrema (maxima and minima) of their temperature perturbation. Thus, the most unstable mode has one maximum only, the second most unstable mode has one maximum and one minimum, and so on. The position of the largest extremum is shifted toward larger values of $r/R$ as the order of the mode increases. In addition, the growth rate (indicated at the top of the panels of Fig.~\ref{fig:eigen1}) decreases with the order of the mode. These two results are represented together in Figure~\ref{fig:growth20}, which displays the growth rate of the  20 most unstable modes as a function of the position of their largest extremum. In comparison with the thermal continuum, slightly smaller values of the growth rate are obtained for the quasi-continuum modes. The displacement of the largest extremum of the discrete modes toward larger values of $r/R$ as their growth rate decreases is consistent with the behavior of the thermal continuum. From  Figure~\ref{fig:growth20} we also see that the solutions are closer to each other as their order increases. This result can be explained by taking into account that the perpendicular thermal conductivity (Equation~(\ref{eq:ionskappa})) is $\kappa_{\perp} \sim \rho_0^2 T^{-1/2}_0$. As we move toward larger $r/R$, $T_0$ increases and $\rho_0$ decreases in the equilibrium, meaning that $\kappa_{\perp}$ gets smaller. Thus, the characteristic perpendicular spatial-scale decreases, causing the modes to cluster as $r/R$  increases. 

\begin{figure}[!htp]
\centering
 \epsscale{0.49}
 \plotone{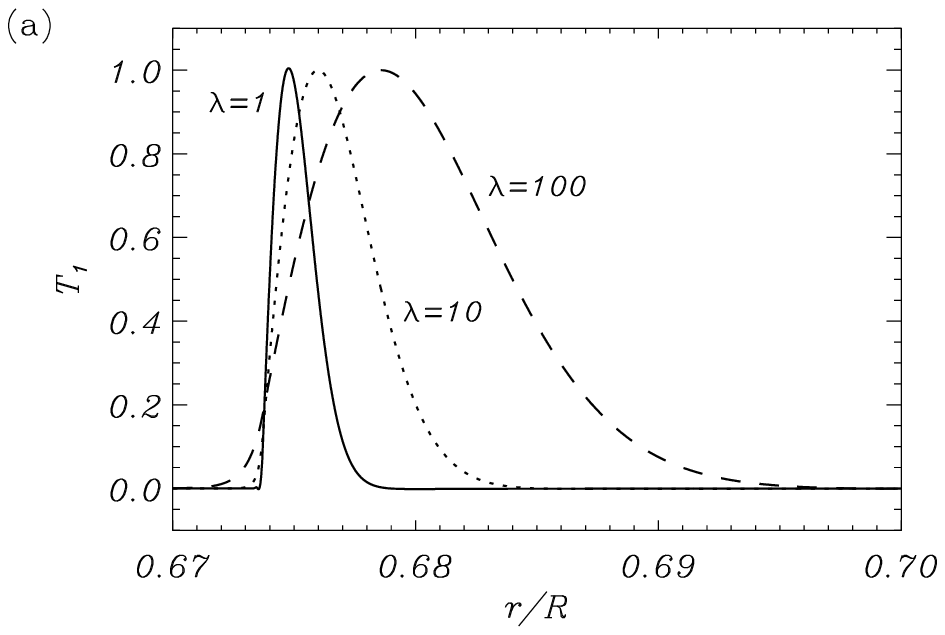}
 \plotone{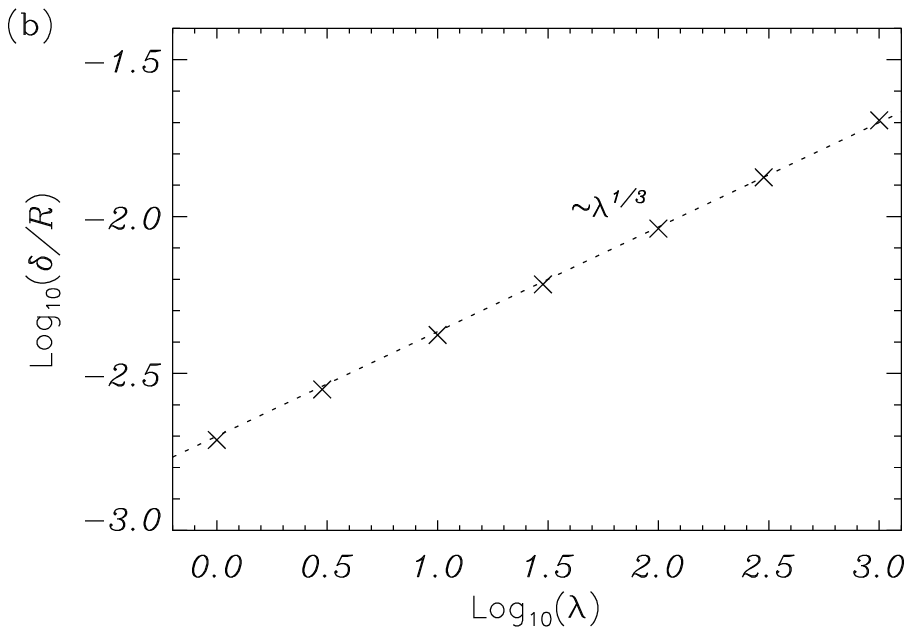}
\caption{(a) Temperature perturbation (in arbitrary units) of the most unstable mode. The different line styles correspond to different values of the perpendicular conductivity enhancing factor: $\lambda = 1$ (solid), $\lambda = 10$ (dotted), and $\lambda = 100$ (dashed).  (b) Width of the maximum of the normalized temperature perturbation, $\delta/R$, corresponding to the most unstable mode (symbol $\times$) versus the perpendicular conductivity enhancing factor $\lambda$. The dotted line corresponds to the dependence $\delta/R \sim \lambda^{1/3}$. Results are computed for $\eta = 0$, $k_z R = 10^{-1}$, and $m=0$. The sinusoidal temperature profile with $l/R=1$ has been used.  \label{fig:growth202}}
\end{figure}

To shed more light on the behavior of the eigenfunctions with $\kappa_\perp$, we perform the substitution $\kappa_\perp  \to \lambda \kappa_\perp$ in the basic equations, with $\lambda$ an enhancing factor. We can artificially increase the value of  $\kappa_\perp$ by means of $\lambda$ and assess its influence on the eigenfunctions. We restrict ourselves to the most unstable mode. Figure~\ref{fig:growth202}(a) shows the temperature eigenfunction of the most unstable mode for different values of $\lambda$. As $\lambda$ grows, the temperature eigenfunction becomes broader. To quantify this effect, we compute the width $\delta$ of the maximum of the temperature perturbation measured at its half height. The parameter $\delta$ is related to the thickness of the conductive layer. Following the method by \citet{SGH91} originally used for Alfv\'en and slow resonances, \citet{vanderlinden93} obtained that the thickness of the conductive layer is proportional to $\kappa_\perp^{1/3}$. In order to compare the analytical result of \citet{vanderlinden93} with our numerical computations, Figure~\ref{fig:growth202}(b) shows $\delta/R$ versus the enhancing parameter $\lambda$. We see that the dependence of $\delta/R$ with $\lambda$ is consistent with a scaling law of $\lambda^{1/3}$, pointing out that our numerical code works properly and recovers the behavior analytically predicted by \citet{vanderlinden93}.

\begin{figure}[!htp]
\centering
 \epsscale{0.49}
 \plotone{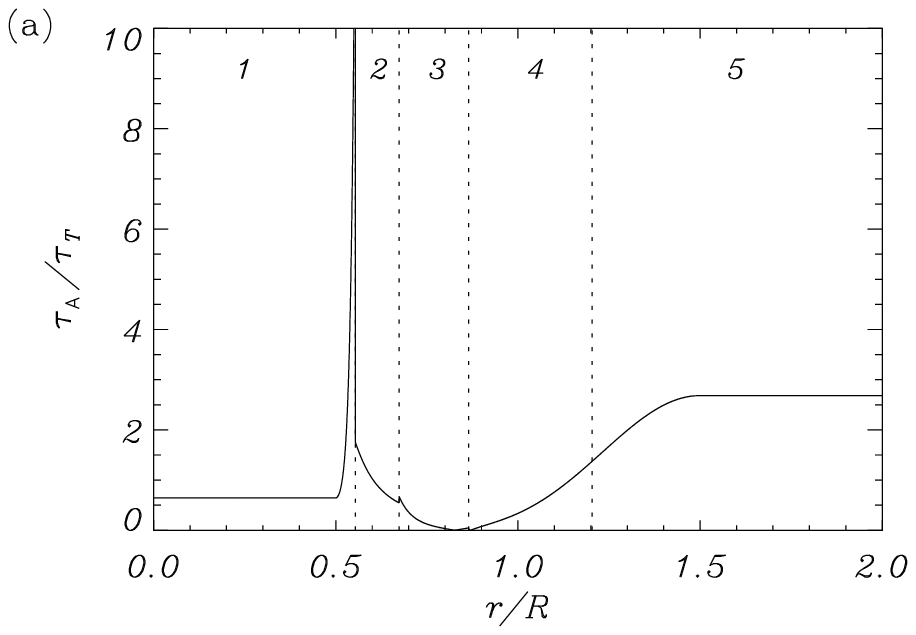}
 \plotone{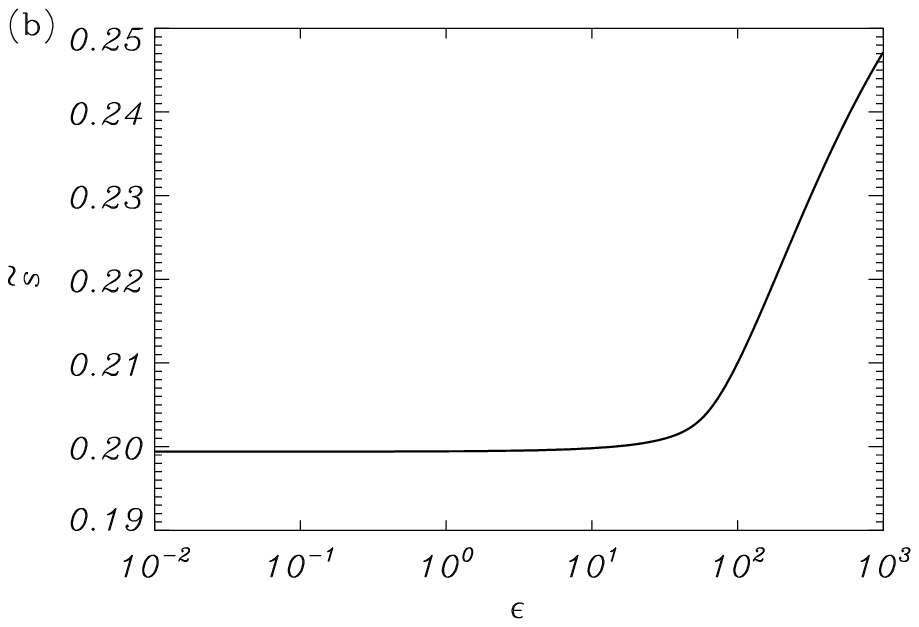}
\caption{(a) Ratio of the Alfv\'en time scale, $\tau_{\rm A}$, to the thermal continuum time scale, $\tau_T$. The vertical dotted lines denote the boundaries between the different parametrizations of the radiation function according to Table~\ref{tab:regimes}. (b) Normalized growth rate of the most unstable mode versus the diffusivity enhancing factor $\epsilon$. The sinusoidal temperature profile with $l/R=1$ has been used. \label{fig:condition}}
\end{figure}


We have also computed the eigenfunctions for other values of the azimuthal wavenumber $m$. We do not have obtained variations of the spectrum of quasi-continuum modes for different values of $m$.  It is worth noting that for $m = 0$, the perturbations $v_\varphi$ and $B_\varphi$ are decoupled from the remaining perturbations, while for $m \ne 0$ all perturbations are coupled. The temperature perturbation is independent of $m$. The effect of $m$ on the growth rate is studied in Section~\ref{sec:param}.

\subsection{Effect of magnetic diffusion}

Here, we first consider magnetic diffusion and neglect perpendicular thermal conduction, i.e., $\kappa_\perp = 0$ and $\eta \ne 0$. \citet{ireland92,ireland98} studied the effect of magnetic diffusion on the thermal continuum. For their particular equilibrium, these authors studied two limit cases of the relative values of the thermal and Alfv\'en time scales. When the thermal time scale is much longer than the Alfv\'en time scale, \citet{ireland92,ireland98} found that the effect of diffusion is to replace the thermal continuum by a dense set of discrete modes, whose growth rate is displaced with respect to the values of the thermal continuum. The displacement of the growth rate is independent of the value of the diffusivity. This is the case of the {\em cool profile} studied by  \citet{ireland92,ireland98}. On the other hand, if the thermal time scale is much shorter than the Alfv\'en time scale, the thermal continuum is not affected by diffusion and the discrete modes are absent. This other possibility corresponds to the {\em hot profile} of  \citet{ireland92,ireland98}. 

To assess the effect of magnetic diffusion on the thermal continuum of our equilibrium, let us compare the Alfv\'en time scale, $\tau_{\rm A} = L/\va $, with the thermal time scale, $\tau_T = 1/|s|$, where $s$ is here the thermal continuum growth rate and $L$ is a typical length scale. We relate $L$ with the longitudinal wavelength of the perturbation, namely $L=2 \pi / k_z$. This is done in Figure~\ref{fig:condition}(a), which displays the ratio $\tau_{\rm A} / \tau_T$ versus $r/R$ with $k_z R = 10^{-1}$. We obtain that in the most unstable part of the continuum, i.e., the beginning of zone 3, $\tau_T$ and $\tau_{\rm A}$ are of the same order. Therefore, we are in a situation between the limit cases studied by \citet{ireland92,ireland98}. We have numerically checked that the growth rates are not modified by the presence of magnetic diffusion. In addition, by setting $\eta \neq 0$ and $\kappa_\perp = 0$ in the equations implemented in the PDE2D code, we have found no discrete thermal solutions. Therefore, the behavior of our equilibrium is similar to that of the {\em hot profile} of \citet{ireland92,ireland98} although in our case $\tau_T$ and $\tau_{\rm A}$ are of the same order.

\begin{figure*}[!htp]
\centering
 \epsscale{0.99}
 \plotone{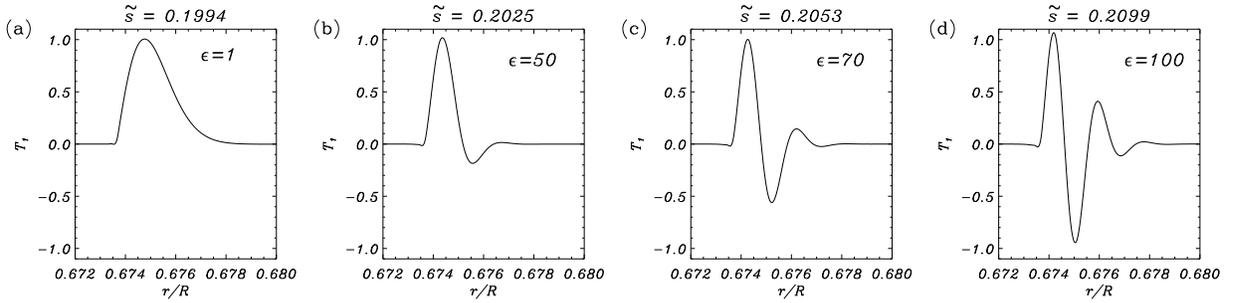}
\caption{Temperature perturbation (in arbitrary units) of the most unstable mode for a diffusivity enhancing factor of (a) $\epsilon=1$, (b) $\epsilon=50$, (c) $\epsilon=70$, and (d) $\epsilon=100$. Computations performed considering the sinusoidal temperature profile with $l/R=1$, $k_z R = 10^{-1}$, and $m=0$. The corresponding normalized growth rate, $\tilde{s}$, is displayed on top of each panel. \label{fig:eigeneta}}
\end{figure*}

A different issue is to determine how the previously described modes for $\kappa_\perp \ne 0$ and $\eta = 0$ (Sec~\ref{sec:kappa}) are affected when magnetic diffusion is included. Hence, we consider the general case $\kappa_\perp \ne 0$ and $\eta \ne 0$. As was done for the perpendicular thermal conduction, we perform the substitution $\eta \to \epsilon \eta$ in the basic equations, with $\epsilon$ an enhancing factor. We can artificially increase the value of  $\eta$ by means of $\epsilon$ and assess its influence. As before, we restrict ourselves to the most unstable  mode, whose growth rate versus $\epsilon$ is displayed in Figure~\ref{fig:condition}(b). For $\epsilon \lesssim 50$ the growth rate is constant and coincides with the value for $\eta = 0$, while the growth rate increases for $\epsilon \gtrsim 50$. For such large values of $\epsilon$, magnetic diffusion governs the behavior of the solutions. When the actual value of $\eta$ is considered, i.e., $\epsilon=1$, the growth rate is approximately the same as that obtained in the absence of magnetic diffusion. Thus, we need much larger, unrealistic values of $\eta$ for the growth rate to be affected by diffusion. On the other hand, Figure~\ref{fig:eigeneta} shows the evolution of the temperature perturbation of the most unstable mode as $\epsilon$ increases. Again, the result for $\epsilon = 1$ is almost indistinguishable from the eigenfunction for $\eta = 0$ (compare Figs.~\ref{fig:eigen1}(a) and \ref{fig:eigeneta}(a)). When $\epsilon$ is increased and so very large, unrealistic values of $\eta$ are considered, the eigenfunction is affected by an additional modulation caused by magnetic diffusion and develops smaller spatial scales as $\epsilon$ is increased. On the basis of these results, we conclude that, for realistic values of the diffusivity, magnetic diffusion is irrelevant for both the growth rate and the eigenfunctions of the most unstable modes.

\begin{figure}[!t]
\centering
 \epsscale{0.49}
 \plotone{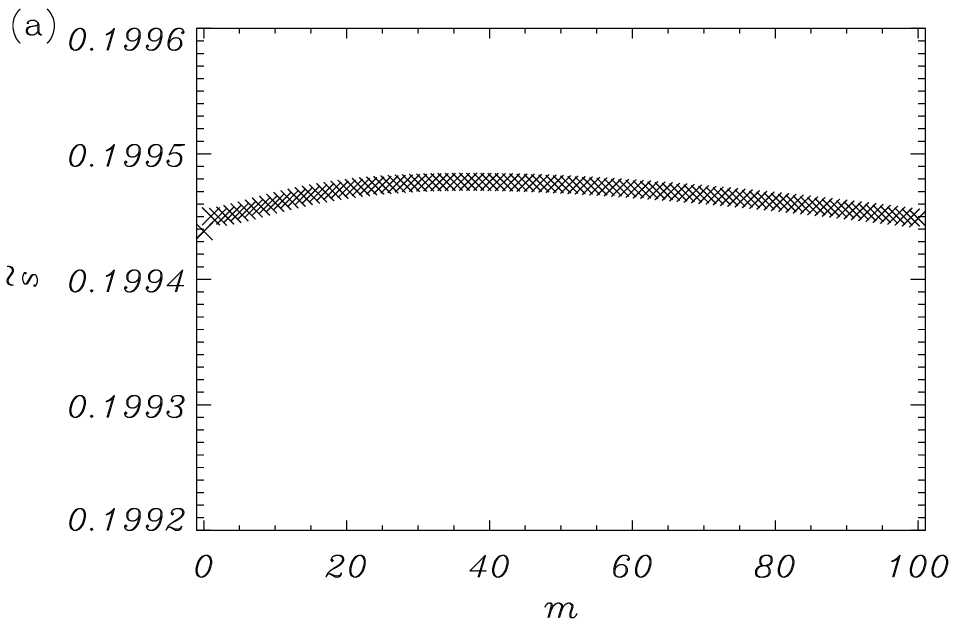}
 \plotone{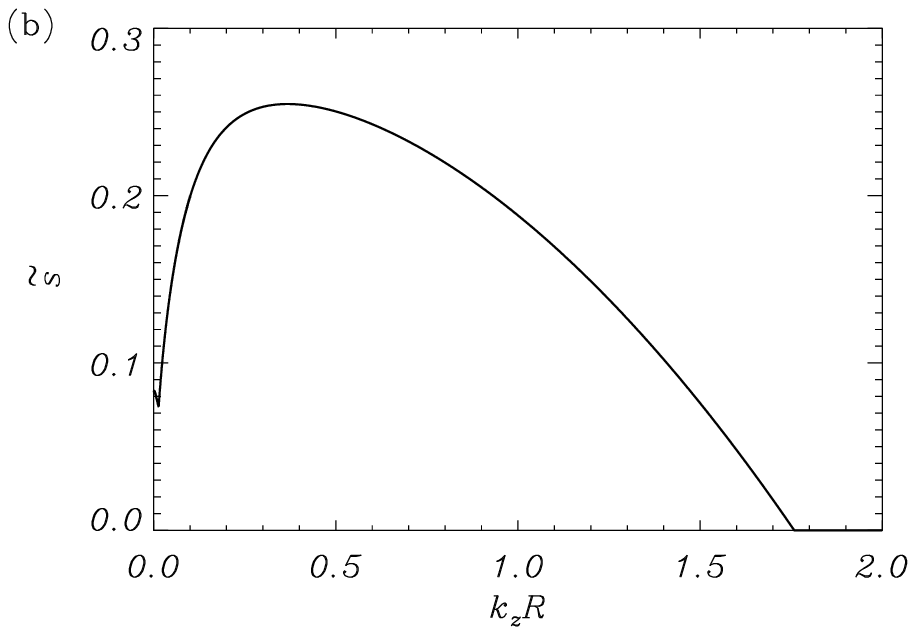}
\caption{Normalized growth rate of the most unstable mode as a function of (a) the azimuthal wavenumber, $m$, for $k_z R = 10^{-1}$, and (b) the dimensionless longitudinal wavenumber, $k_z R$, for $m=0$. The results of both panels correspond to the sinusoidal temperature profile with $l/R=1$. \label{fig:mazilayer}}
\end{figure}

\begin{figure}[!htp]
\centering
 \epsscale{0.49}
 \plotone{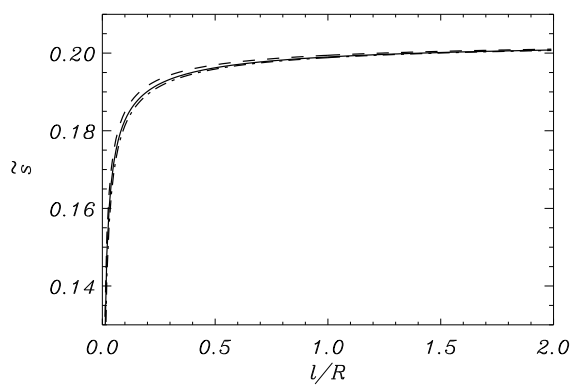}
\caption{Normalized growth rate of the most unstable mode versus the thickness of the transitional layer, $l/R$. The different line styles in panel, almost superimposed, represent the linear (solid), sinusoidal (dashed), and Gaussian (dot-dashed) temperature profiles.  All computations are performed with $k_z R = 10^{-1}$ and $m=0$. \label{fig:kz}}
\end{figure}

\subsection{Comparison between the Hildner and Klimchuk-Raymond radiative loss functions}

We have performed some test computations using Klimchuk-Raymond's fit for the radiative loss function in order to compare with the results obtained using Hildner's fit (discussed in the previous Subsections). Regarding the form of the temperature perturbations, we find no significant differences. The maximum of the temperature perturbation is shifted toward slightly larger values of $r$ for Klimchuk-Raymond's fit. For the fundamental mode, the maximum of $T_1$ takes place at $r/R \approx 0.675$ using Hildner's fit, and at $r/R \approx 0.69$ using Klimchuk-Raymond's fit for $k_z R = 10^{-1}$, $l/R = 1$, and $m=0$. On the other hand, the growth rate is larger for Klimchuk-Raymond's parametrization. This is in agreement with the behavior of the thermal continuum (see Figs.~\ref{fig:continuum}(a) and \ref{fig:continuumklim}).  For the same set of parameters as before, the growth rate of the most unstable mode is $\tilde{s} \approx 0.1994$ using Hildner's fit, and $\tilde{s} \approx 0.4370$ using Klimchuk-Raymond's fit. Apart from this difference in the growth rate, the results for both Hildner's and Klimchuk-Raymond's parametrizations are equivalent.

\section{PARAMETRIC STUDY OF THE GROWTH RATES}

\label{sec:param}

In this Section, we study the growth rate of the most unstable mode as a function of the dimensionless longitudinal wavenumber, $k_z R$, the azimuthal wavenumber, $m$, and the thickness of the transitional layer, $l/R$. Hildner's fit for the radiative loss function is used in all cases.

First, we assess the effect of the azimuthal wavenumber, $m$, (see Fig.~\ref{fig:mazilayer}(a)). Note that only positive integer values of $m$ are considered in Figure~\ref{fig:mazilayer}(a) as the results are equivalent when negative values are used. As for the thermal continuum, the value of $m$ is irrelevant for the growth rate of the discrete thermal modes. The almost negligible variation of the growth rate with $m$ shown in Figure~\ref{fig:mazilayer}(a) might not be physical and might be attributed to the accuracy of the numerical code. On the other hand, the dependence on the longitudinal wavenumber is displayed in Figure~\ref{fig:mazilayer}(b). The maximum of the growth rate takes place for $k_z R \approx 0.4$, while $\tilde{s}$ decreases as $k_z R$ is increased or reduced. For $k_z R \gtrsim 0.4$, the growth rate decreases and the thermal mode is stabilized by the effect of thermal conduction parallel to magnetic field lines,  as commented in Section~\ref{sec:therm}. The critical wavenumber for the stabilization is consistent with the expression given in Equation~(\ref{eq:kzcrit}) for the stabilization of the thermal continuum. The growth rate also decreases with respect to the maximum value when $k_z R \lesssim 0.4$. This behavior is different from the behavior of the thermal continuum (compare Figs.~\ref{fig:continuum2} and \ref{fig:mazilayer}(b)).  In addition, we find that for $k_z R \to 0$ a rich collection of couplings between the different thermal modes takes place. This complex system of couplings is not visible on the scale of Figure~\ref{fig:mazilayer}(b). We do not explore these couplings in detail here because they take place for very small $k_z R$ and far from the maximum value of the growth rate.

Finally, the dependence on the thickness of the transitional layer, $l/R$, is shown on Figure~\ref{fig:kz}. In a real prominence thread, the thickness of the transitional layer is probably determined by the coefficient of thermal conduction perpendicular to the magnetic field. In fully ionized coronal plasmas, the perpendicular thermal conductivity is very small. However, for cool prominence temperatures the plasma is only partially ionized and the perpendicular thermal conductivity increases by several orders of magnitude due to the effect of thermal conduction by neutrals \citep[see, e.g.,][]{parker,ibanez,forteza08}. This effect might be important in the inner part of the PCTR. Unfortunately, since the ionization degree in prominence threads is unknown \citep[see recent estimations][]{labrosse}, the thickness of the transitional layer cannot be determined with confidence. For this reason, we take $l/R$  as a free parameter. The growth rate is almost constant for $l/R \gtrsim 0.5$, whereas $\tilde{s}$ decreases for  $l/R \lesssim 0.5$. When $l/R \to 0$, $\tilde{s} \to 0$ because the transitional layer is absent and the thermal continuum disappears. As in previous computations, the results for the different temperature profiles are almost identical, hence the form of the temperature profile is not relevant for the growth rate. This is an important result because it means that our present conclusions apply for all possible temperature profiles in the particular PCTR of prominence threads. 

\section{IMPLICATIONS FOR THREAD LIFETIMES}

\label{sec:discussion}

In this paper we have shown that prominence threads, modeled as cool and dense magnetic flux tubes embedded in a much hotter and less dense coronal environment, are not stable because unstable thermal modes are present in their particular PCTR. This statement qualitatively agrees with the results from \citet{karpen}, who studied thermal instabilities in a two-dimensional spatially random magnetic field distribution, and concluded that cool condensations cannot be in static equilibrium with a hot exterior. However, in the work by \citet{karpen} the surrounding of the condensations simply continue to cool down because these authors did not consider the effect of thermal conduction perpendicular to the magnetic field.  The unstable discrete modes present in our model appear due to the effect of thermal conduction perpendicular to the magnetic field. Our results for the growth rate of the unstable thermal modes have direct implications for the stability and lifetime of prominence threads. In our analysis, we have restricted ourselves to the linear stage of the thermal instability. This stage represents the initial phase in which the unstable thermal mode grows after its excitation.  The combined influence of the different unstable modes will affect the subsequent nonlinear evolution of the plasma within the PCTR, because many of the unstable modes have very similar growth rates and the evolution will not be dominated by a single mode.  Significant changes in the equilibrium of the threads would take place during the subsequent nonlinear evolution of the thermal instability, because of the large temperature and density gradients generated, which  may be detected by the observations. The growth rate of the linear phase provides us with a reasonable estimation of the typical time scale on which the effect of the thermal instability in the PCTR of the threads would be observable and may affect the thread dynamics. 

\begin{figure}[!htp]
\centering
 \epsscale{0.49}
 \plotone{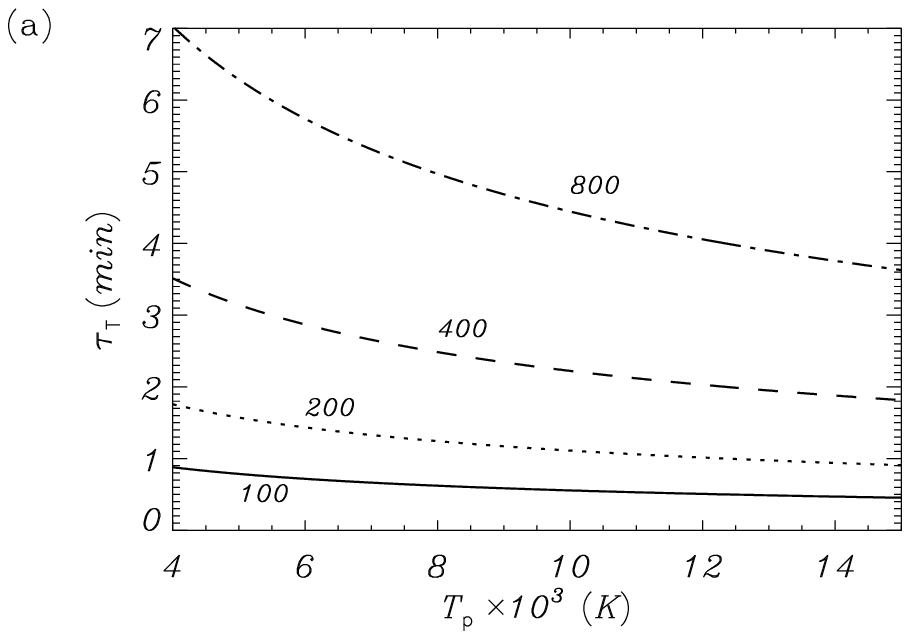} 
\plotone{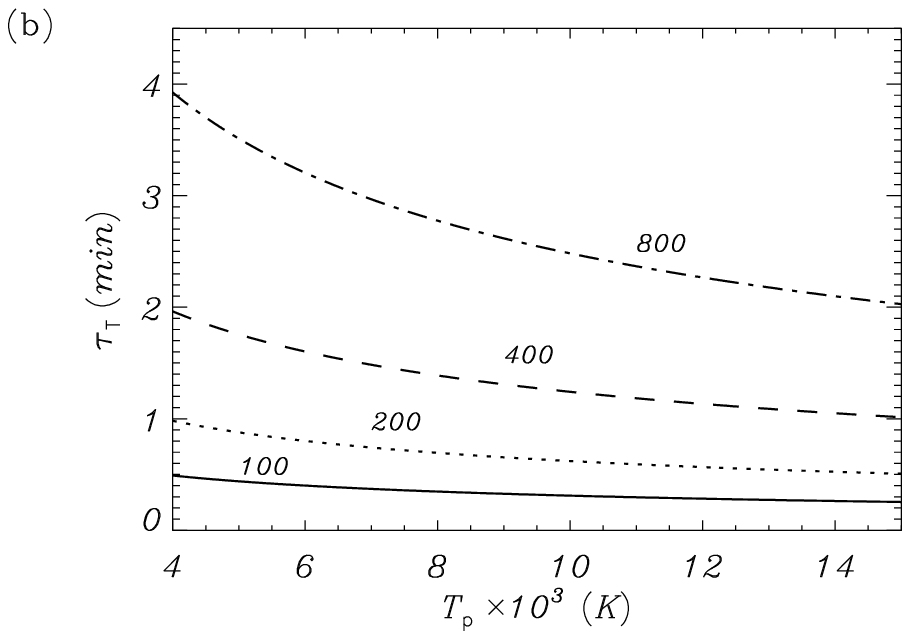}
\caption{Thermal instability time scale, $\tau_T$, computed from the maximum growth of the most unstable mode versus the temperature of the thread core, $T_{\rm p}$, using (a) Hildner's and (b) Klimchuk-Raymond's parametrization for the radiative loss function. The different line styles in both panels represent different values of the thread radius: $R=$~100~km (solid), 200~km (dotted), 400~km (dashed), and 800~km (dot-dashed).  \label{fig:times}}
\end{figure}

 In the analysis of the previous Sections, we have seen that the most relevant parameter affecting the instability growth rate is the longitudinal wavenumber, since it determines the efficiency of parallel thermal conduction as stabilizing mechanism. If unstable thermal modes are excited in a prominence thread by an arbitrary disturbance, the instability would be dominated by the most unstable solution. We find that in our model the maximum of the growth takes place for $k_z R \approx 0.4$, and corresponds to a dimensionless growth rate of $\tilde{s} \approx 0.255$ using Hildner's radiative loss function, and $\tilde{s} \approx 0.467$ using Klimchuk-Raymond's radiative loss function. The corresponding instability time scale is $\tau_T = 1/s$, with $s$ the dimensional growth rate computed from its dimensionless value as $s=\tilde{s} \csp / R$. In Figure~\ref{fig:times}, we plot $\tau_T$ as a function of the temperature at the center of the thread, $T_{\rm p}$, for different values of the thread radius, $R$. We find that, for realistic values of $T_{\rm p}$ and $R$, the instability time scale is of the order of a few minutes. We obtain that  $\tau_T$ decreases as $T_{\rm p}$ gets higher and increases as $R$ gets larger, meaning that cool and wide threads are more thermally stable than hot and thin threads. We also see that Klimchuk-Raymond's parametrization produces smaller instability time scales than with Hildner's parametrization.  The values of $\tau_T$ indicated in Figure~\ref{fig:times} are of the same order of magnitude as the typically observed lasting time of the threads in sequences of H$\alpha$ images of solar filaments \citep[e.g.,][]{lin04,lin05,lin09}. This result implies that the thermal instability may play a relevant role for the dynamics and stability of prominence and filament fine structures, because the time scale related to the thermal instability is consistent with the observed time scale on which threads seem to suffer important changes in their equilibrium.

\section{DISCUSSION AND SUMMARY}
\label{sec:sum}

In this paper we have studied the properties of quasi-continuum thermal modes in a prominence thread model which is transversely inhomogeneous. We have followed the method of \citet{vanderlindencont} and \citet{vanderlinden91} to investigate, first, the stability of the thermal continuum and, later, the effect of cross-field thermal conduction and magnetic diffusion. We have recovered the general results by \citet{vanderlindencont} and \citet{vanderlinden91} in all cases studied in the present paper. In particular, we have found that the thermal continuum in prominence threads is unstable for PCTR temperatures. In agreement with the results of \citet{vanderlinden91}, the effect of cross-field thermal conduction is to replace the thermal continuum by discrete modes which retain the unstable character of the continuum. On the contrary, the role of magnetic diffusion is negligible in our model \citep[see][]{ireland92,ireland98}.  We have obtained the important result that the instability growth rate of the most unstable mode is independent of the form of the temperature profile within the PCTR of the thread, and the instability time scale is consistent with the observed lifetime of the threads in H$\alpha$ observations of solar filaments \citep[e.g.,][]{lin08,lin09}. Considering our present results along with those obtained by \citet{vanderlinden91} and \citet{vanderlinden93}, we conclude that unstable thermal modes may play a relevant role for both the formation of the prominence fine structure and the subsequent instability of the thin prominence threads. 

Here we have considered a fully ionized prominence. However, due to the low temperature in the dense core of the threads, the cool prominence plasma is expected to be partially ionized \citep{labrosse}. If partial ionization is taken into account, several additional effects have to be considered, namely thermal conduction by neutrals, ambipolar (or Cowling's) diffusion caused by ion-neutral collisions, and a modified radiation function \citep[see, e.g.,][]{forteza08,solerhelium}. The role of both effects depends strongly on the plasma ionization degree. It requires complicated computations of NLTE radiative transfer and statistical equilibrium of atomic level populations to obtain the precise profile of the ionization degree in prominence threads \citep[see extensive details in, e.g.,][]{labrosse,labrossereview}. Nevertheless, some relevant conclusions can be obtained if one reasonably assumes that the ionization degree in prominence plasmas is mainly determined by the temperature and, to a lesser extend, by the density.  While the ionization degree of the core of the thread is unknown, it is realistic to assume that the condition of full ionization occurs in the PCTR for temperatures higher than a critical temperature, namely $T^*$. The results by \citet{labrosse} and \citet{schure} suggest realistic values for the critical ionizing temperature for the hydrogen plasma in prominences of $T^* \approx 2 \times 10^4$~K and $T^* \approx 2.5 \times 10^4$~K, respectively, so only the inner part of the transitional layer would be partially ionized. According the instability criterion (Equation~(\ref{eq:crit1})), the unstable part of the thermal continuum takes place at temperatures higher than $8 \times 10^4$~K, which is higher than the critical ionizing temperature. This suggests that the ionization degree of the cool part of the fine structure is probably not relevant for the unstable quasi-continuum modes, because they are mainly confined where hydrogen may be already fully ionized. However, a detailed investigation of this issue is needed for more robust conclusions.

The present investigation is a first step and should be improved in the future by taking into account relevant effects that have not been included in our model. Among these effects, the role of density and magnetic longitudinal inhomogeneity and the effect of magnetic twist will be assessed in forthcoming works. We have also neglected the effect of mass flows. Previous studies on linear thermal modes in flowing homogeneous prominence plasmas \citep[see, e.g.,][]{solernonad,carbonell09, carbonell10, barcelo11} suggests that flows along magnetic field lines do not affect the linear growth rate of the thermal instability.

\acknowledgements{
   We thank Ronald Van der Linden for reading the manuscript and for giving helpful comments. This work was started when RS was a PhD student in UIB. RS is grateful to the CAIB for financial support. RS acknowledges support from a postdoctoral fellowship in Leuven within the EU Research and Training Network SOLAIRE (MTRN-CT-2006-035484). RS and JLB acknowledges the financial support received from the Spanish MICINN and FEDER funds (AYA2006-07637).  RS and JLB also acknowledge discussion within ISSI Team on Solar Prominence Formation and Equilibrium: New data, new models. MG acknowledges support from K.U. Leuven via GOA/2009-009.}


\begin{thebibliography}{}

\bibitem[Appert et al.(1974)]{appert} Appert, K, Gruber, R., \& Vaclavik, J. 1974, Phys. Fluids, 17, 1471

%
%
%
 \bibitem[Arregui \& Ballester(2010)]{arreguiballester} Arregui, I., \& Ballester, J. L. 2010, \ssr, in press
%
%
 \bibitem[Ballester \& Priest(1989)]{ballesterpriest} Ballester, J. L., \& Priest, E. R. 1989, \aap, 225, 213
%
%
%
 \bibitem[Ballester(2006)]{ballester} Ballester, J. L. 2006, Phil. Trans. R. Soc. A, 364, 405 
%

 \bibitem[Barcel\'o et al.(2011)]{barcelo11} Barcel\'o, S., Carbonell, M., \& Ballester, J. L. 2011, \aap, 525, A60 

%
 \bibitem[Berger et al.(2008)]{berger} Berger et al. 2008, \apj, 676, L89

\bibitem[Braginskii(1965)]{brag} Braginskii, S. I. 1965, Rev. Plasma Phys., 1, 205

   \bibitem[Carbonell et al.(2004)]{carbonell04} Carbonell, M., Oliver, R., \& Ballester, J. L. 2004, \aap, 415, 739 

\bibitem[Carbonell et al.(2009)]{carbonell09} Carbonell, M., Oliver, R., \& Ballester, J. L. 2009, NewA, 14, 277 

\bibitem[Carbonell et al.(2010)]{carbonell10} Carbonell, M., Forteza, P., Oliver, R., \& Ballester, J. L. 2010, \aap, 515,  A80


 \bibitem[Chae et al.(2008)]{chae} Chae, J., Ahn, K., Lim, E.-K., Choe, G. S., \& Sakurai, T. 2008, \apj, 689, L73

 \bibitem[Chae et al.(2010)]{chae2} Chae, J. 2010, \apj, 714, 618	

\bibitem[Cirigliano et al.(2004)]{cirigliano} Cirigliano, D., Vial, J.-C., Rovira, M. 2004, \solphys, 223, 95

\bibitem[Cox \& Tucker(1969)]{coxtucker} Cox, D. P., \& Tucker, W. H. 1969, \apj, 157, 1157


%
%
%
%
%
%

\bibitem[Engvold(1976)]{engvold76} Engvold, O. 1976, \solphys, 49, 283

\bibitem[Engvold(2004)]{engvold2004} Engvold, O. 2004, in IAU Symp. 223, Multi-Wavelength Investigations of Solar Activity, ed. A. V. Stepanov, E. E. Benevolenskaya, and A. G. Kosovichev (Cambridge: Cambridge Univ. Press), 187

  \bibitem[Engvold(2008)]{engvold} Engvold, O. 2008, in IAU Symp. 247, Waves \& Oscillations in the Solar Atmosphere: Heating and Magneto-Seismology, ed. R. Erd\'elyi \& C. A. Mendoza-Brice\~no (Cambridge: Cambridge Univ. Press), 152

\bibitem[Field(1965)]{field} Field, G. B. 1965, \apj, 142, 531

%
  \bibitem[Forteza et al.(2008)]{forteza08} Forteza, P., Oliver, R., \& Ballester, J. L. 2008, \aap, 492, 223



\bibitem[Gouttebroze \& Labrosse(2009)]{labrosse} Gouttebroze, P., \& Labrosse, N. 2009, \aap, 503, 663

\bibitem[Heinzel(2007)]{heinzel} Heinzel, P. 2007, ASPC, 368, 271

\bibitem[Heyvaerts(1974)]{heyvaerts} Heyvaerts, J. 1974, \aap, 34, 65

   \bibitem[Hildner(1974)]{hildner} Hildner, E. 1974, \solphys, 35, 123



\bibitem[Ib\'a\~nez \& Mendoza(1990)]{ibanez} Ib\'a\~nez, M. H., \& Mendoza, C. A. 1990, \apss, 164, 193

\bibitem[Ireland et al.(1992)]{ireland92} Ireland, R. C., Van der Linden, R. A. M., Hood, A. W., \& Goossens, M. 1992, \solphys, 142, 265

\bibitem[Ireland et al.(1998)]{ireland98} Ireland, R. C., Van der Linden, R. A. M., \& Hood, A. W. 1998, \solphys, 179, 115

%

\bibitem[Karpen et al.(1989)]{karpen} Karpen, J. T., Antiochos, S. K., Picone, J. M., Dahlburg, R. B. 1989, \apj, 338, 493

\bibitem[Klimchuk \& Cargill(2001)]{klimchuk} Klimchuk, J. A., \& Cargill, P. J. 2001, \apj, 553, 440

\bibitem[Labrosse et al.(2010)]{labrossereview} Labrosse, N., Heinzel, P., Vial, J.-C., Kucera, T., Parenti, S., Gun\'ar, S., Schmieder, B., \& Kilper, G. 2010, \ssr, 151, 243

    \bibitem[Lin(2004)]{lin04} Lin, Y. 2004, PhD Thesis, University of Oslo, Norway
%

 \bibitem[Lin et al.(2003)]{lin03} Lin, Y., Engvold, O., \& Wiik, J. E. 2003, \solphys, 216, 109

%
     \bibitem[Lin et al.(2005)]{lin05} Lin, Y.,  Engvold, O., Rouppe van der Voort, L. H. M., Wiik, J. E., \& Berger, T. E. 2005, \solphys, 226, 239


    \bibitem[Lin et al.(2007)]{lin07} Lin, Y., Engvold, O., Rouppe van der Voort, L. H. M., \& van Noort, M.  2007, \solphys, 246, 65
%
  \bibitem[Lin et al.(2008)]{lin08} Lin, Y., Martin, S. F., \& Engvold, O. 2008, in ASP Conf. Ser. 383, Subsurface and Atmospheric Influences on Solar Activity, ed. R. Howe, R. W. Komm, K. S. Balasubramaniam, \& G. J. D. Petrie (San Francisco: ASP), 235
%
%
 \bibitem[Lin et al.(2009)]{lin09} Lin, Y., Soler, R., Engvold, O., Ballester, J. L., Langangen, \O., Oliver, R., \& Rouppe van der Voort, L. H. M. 2009, \apj, 704, 870
%
%
   \bibitem[Mackay et al.(2010)]{mackay} Mackay, D. H., Karpen, J. T., Ballester, J. L., Schmieder, B., \& Aulanier, G. 2010, \ssr, 151, 333

\bibitem[Menzel \& Wolbach(1960)]{menzel60} Menzel, D. H., \& Wolbach, J. G. 1960, \aj, 65, 54

\bibitem[Milne et al.(1979)]{milne} Milne, A. M., Priest, E. R., \& Roberts, B. 1979, \apj, 232, 304

%
 \bibitem[Ning et al.(2009)]{ning} Ning, Z., Cao, W., Okamoto, T. J., Ichimoto, K., \& Qu, Z. Q. 2009, \aap, 499, 595
%
%
  \bibitem[Okamoto et al.(2007)]{okamoto} Okamoto, T. J, et al. 2007, Science, 318, 1557 
%
 \bibitem[Oliver(2009)]{oliver} Oliver, R. 2009, \ssr, 149, 175

  \bibitem[Parker(1953)]{parker} Parker, E. N. 1953, \apj, 117, 431


\bibitem[Poedts \& Kerner(1991)]{stefaan} Poedts, S., \& Kerner, W. 1991, Phys. Rev. Let., 66, 2871

\bibitem[Priest(1982)]{priest} Priest, E. R. 1982, Solar magnetohydrodynamics, D. Reitel Publishing Company
%
 \bibitem[Rempel et al.(1999)]{rempel} Rempel, M., Schmitt, D., \& Glatzel, W. 1999, \aap, 343, 615


  \bibitem[Rosner et al.(1978)]{rosner} Rosner, R., Tucker, W. H. \& Vaiana, G. S. 1978, \apj, 220, 643

%
 \bibitem[Sakurai et al.(1991)]{SGH91} Sakurai, T., Goossens, M, \& Hollweg, J. V. 1991, \solphys, 133, 227
%

\bibitem[Schmieder et al.(2010)]{brigi} Schmieder, B., Chandra, R., Berlicki, A., \& Mein, P. 2010, \aap, in press

\bibitem[Schure et al.(2010)]{schure} Schure, K. M., Kosenko, D., Kaastra, J. S., Keppens, R., \& Vink, J. 2010, \aap, 508, 751

\bibitem[Sewell(2005)]{sewell} Sewell, G. 2005, The Numerical Solution of Ordinary and Partial Differential Equations, Wiley \& Sons

     \bibitem[Soler et al.(2008)]{solernonad} Soler, R., Oliver, R., \& Ballester, J. L. 2008,  \apj, 684, 725 
%
%

  \bibitem[Soler et al.(2010a)]{solerhelium} Soler, R., Oliver, R., \& Ballester, J. L. 2010a, \aap, 512, A28

 \bibitem[Soler et al.(2010b)]{solerKH} Soler, R., Terradas, J., Oliver, R., Ballester, J. L., \& Goossens, M. 2010b,  \apj, 712, 875

\bibitem[Soler(2010)]{solerphd} Soler, R. 2010, PhD Thesis, Universitat de les Illes Balears, Spain

 \bibitem[Spitzer(1962)]{spitzer} Spitzer, L. 1962, Physics of fully ionized gases, Interscience



%



\bibitem[Van der Linden \& Goossens(1991)]{vanderlinden91} Van der Linden, R. A. M., \& Goossens, M. 1991, \solphys, 134, 247

\bibitem[Van der Linden et al.(1991)]{vanderlindencont} Van der Linden, R. A. M., Goossens, M., \& Goedbloed, J. P. 1991, Phys. Fluids B, 3, 866

\bibitem[Van der Linden(1993)]{vanderlinden93} Van der Linden, R. A. M. 1993, Geophys. Astrophys. Fluid Dyn, 69, 183


\bibitem[Zaqarashvili et al.(2010)]{temury}  Zaqarashvili, T. V., D\'iaz, A. J., Oliver, R., \& Ballester, J. L. 2010, \aap, in press

 \bibitem[Zirker et al.(1994)]{zirker94} Zirker, J. B., Engvold, O., \& Yi, Z. 1994, \solphys, 150, 81

 \bibitem[Zirker et al.(1998)]{zirker98} Zirker, J. B., Engvold, O., \& Martin, S. F. 1998, Nature, 396, 440

  \end{thebibliography}
\end{document}